\documentclass[10pt,sort&compress]{article}
\usepackage{graphicx,rotating,multirow}
\usepackage{bm,multirow}
\usepackage{amsmath,amssymb,color,mathrsfs}

\usepackage{footnote}
\usepackage{citesort}
\usepackage{cuted}

\def\lambdabar{\lambda\kern-1ex\raise0.55ex\hbox{--}}

\newcommand{\AJP}{Am. J. Phys. }

\newcommand{\APL}{Appl. Phys. Lett. }
\newcommand{\APB}{Ann. Phys. (Berlin) }
\newcommand{\APNY}{Ann. Phys. (N.Y.) }

\newcommand{\EJP}{Eur. J. Phys. }
\newcommand{\EPJD}{Eur. Phys. J. D }

\newcommand{\JAP}{J. Appl. Phys. }
\newcommand{\JCP}{J. Chem. Phys. }

\newcommand{\JMP}{J. Math. Phys. }
\newcommand{\jpa}{J. Phys. A }
\newcommand{\jpb}{J. Phys. B }
\newcommand{\jpc}{J. Phys. C }
\newcommand{\JPCM}{J. Phys. Condens. Matter }
\newcommand{\LNC}{Lett. Nuovo Cimento }

\newcommand{\NPA}{Nucl. Phys. A }

\newcommand{\PR}{Phys. Rev. }
\newcommand{\PRA}{Phys. Rev. A }
\newcommand{\PRB}{Phys. Rev. B }

\newcommand{\PRE}{Phys. Rev. E }
\newcommand{\PRL}{Phys. Rev. Lett. }
\newcommand{\PRe}{Phys. Rep. }

\newcommand{\PSSB}{Phys. Status Solidi B }

\newcommand{\SPJ}{Sov. Phys. - JETP }
\newcommand{\SSC}{Solid State Commun. }

\newcommand{\ZETF}{Zh. Eksp. Teor. Fiz. }
\newcommand{\ZP}{Z. Phys. }
\newcommand{\ZPA}{Z. Phys. A }
\newcommand{\ZPB}{Z. Phys. B }

\definecolor{officegreen}{rgb}{0,0.5,0}
\definecolor{pakistangreen}{rgb}{0,0.4,0}
\definecolor{palatinatepurple}{rgb}{0.41,0.16,0.38}
\definecolor{sangria}{rgb}{0.57,0,0.04}

\begin{document}
\title{Evolution of electric-field-induced quasibound states and resonances in one-dimensional open quantum systems}
\author{O. Olendski\footnote{Department of Applied Physics and Astronomy, University of Sharjah, P.O. Box 27272, Sharjah, United Arab Emirates; E-mail: oolendski@sharjah.ac.ae}}

\maketitle

\begin{abstract}
A comparative analysis of three different time-independent approaches to studying open quantum structures in a uniform electric field $\mathscr{E}$ was performed using the example of a one-dimensional attractive or repulsive $\delta$-potential and the surface that supports the Robin boundary condition. The three considered methods exploit different properties of the scattering matrix $S(\mathscr{E};E)$ as a function of energy $E$: its poles, real values, and zeros of the second derivative of its phase. The essential feature of the method of zeroing the resolvent, which produces complex energies, is the unlimited growth of the wave function at infinity, which is, however, eliminated by the time-dependent interpretation. The real energies at which the unitary scattering matrix becomes real correspond to the largest possible distortion, $S=+1$, or its absence at $S=-1$ which in  either case leads to the formation of quasibound states. Depending on their response to the increasing electric intensity, two types of field-induced positive energy quasibound levels are identified: electron- and hole-like states. Their evolution and interaction in the enlarging field lead ultimately to the coalescence of pairs of opposite states, with concomitant divergence of the associated dipole moments in what is construed as an electric breakdown of the structure. The characteristic features of the coalescence fields and energies are calculated and the behavior of the levels in their vicinity is analyzed. Similarities between the different approaches and their peculiarities are highlighted; in particular, for the zero-field bound state in the limit of the vanishing $\mathscr{E}$, all three methods produce the same results, with their outcomes deviating from each other according to growing electric intensity. The significance of the zero-field spatial symmetry for the formation, number, and evolution of the electron- and hole-like states, and the interaction between them, is underlined by comparing outcomes for the symmetric $\delta$ geometry and asymmetric Robin wall.
\end{abstract}

\section{Introduction}\label{Introduction}
The most fundamental quantity to analyze in the study of the elastic motion of non-relativistic quantum particles in a potential $V({\bf r})$ that undergoes sufficiently fast decays at infinity is the scattering matrix $S(E)$, which describes the distortion of the field-free wave function by the force exerted on the corpuscle by $V({\bf r})$ \cite{Newton1,Landau1,Baz1,Kukulin1}. $S(E)$ is generally considered to be a dimensionless complex function of the energy $E$ of the particle and, as such, can have poles in the ${\rm Re}(E)-{\rm Im}(E)$ plane, which determine the resonances characterized by (usually) {\em complex} energy. If the potential allows the existence of bound states, the scattering matrix also has poles at their negative energies. Another important property of the matrix $S$ is its unitarity, $|S|=1$, at real energies that do not coincide with the energies of the bound states; this physically expresses the conservation of the number of particles in the elastic collisions. It is natural to expect that for the {\em complex} unitary function $S(E)$ its {\em real} values $\pm1$ can have a special meaning. Additionally, due to the unitarity, the scattering matrix can be expressed in the form
\begin{equation}\label{Unitary1}
S(E)=e^{i\varphi_S},
\end{equation}
with the real phase $\varphi_S$ being a function of the energy, $\varphi_S\equiv\varphi_S(E)$. It is known \cite{Newton1,Kukulin1} that the region where the fastest change of $\varphi_S$ occurs is most significant, and the Wigner delay time $\tau_W$ \cite{Wigner1} is arrived at in this manner. $\tau_W$ is defined as a derivative of the phase $\varphi_S$ with respect to energy
\begin{equation}\label{Wigner1}
\tau_W(E)=\hbar\frac{d\varphi_S}{dE}
\end{equation}
and is an essential characteristic of the scattering process \cite{deCarvalho1,Maquet1}, with the extreme energies at which it achieves its maxima being the most important. Thus, three sets of energy have been identified, all of which are significant for the scattering matrix $S$
\begin{itemize}
\item (in general) {\em complex} energies, at which $S(E)=\infty$,
\item {\em real} energies, at which $S(E)=\pm1$,
\item {\em real} energies, at which $d\tau_W/dE=0$ and $d^2\tau_W/dE^2<0$.
\end{itemize}
Each of these sets defines its own specific type of behavior of $S$ at and around the corresponding energy, and even within each set the physical processes that are mathematically described by the scattering matrix may be quantitatively different. The overwhelming majority of mathematicians prefer to analyze only the first case where, on the basis of the time-independent Schr\"{o}dinger equation, the complex energies are located and calculated for each particular potential without requiring details of the associated waveforms. These functions, as the first stage of the physical consideration reveals, exhibit unrestricted growth at infinity, known as the 'exponential catastrophe' \cite{Bohm1}. However, even more careful interpretation allows it to be eliminated by proper reasoning involving the temporal evolution of the wave function \cite{Baz1,Bohm1,Holstein1}. The real part of the complex energy is customarily associated with the location of the resonance on the $E$ axis, whereas its imaginary component describes its half width or lifetime. For each specific potential, physicists also analyze the last two energy sets, but quite often this is performed separately for each case and no parallels are drawn between them and the first (complex energy) counterpart.

In the present study, a comparative analysis was performed of the outcomes of the three above-mentioned approaches, as applied to the behavior of electrons in the uniform electric field $\mathscr{E}$ superimposed on (a) a one-dimensional (1D) attractive or repulsive $\delta$-potential, which is the extreme limit of the finite width and depth quantum well (QW) or finite height quantum barrier, and (b) the Robin wall, i.e., a surface ${\cal S}$ that in the general 3D geometry supports the boundary condition (BC) for the wave function $\Psi(\bf r)$ of the form\cite{Gustafson1}
\begin{equation}\label{Robin1}
\left.{\bf n}{\bm\nabla}\Psi\right|_{\cal S}=\left.\frac{1}{\Lambda}\Psi\right|_{\cal S},
\end{equation}
where $\bf n$ is a unit inward vector and the extrapolation length $\Lambda$, which is considered to be real, is called the Robin or de Gennes \cite{deGennes1} distance. The model that represents an extremely localized finite strength interaction in the form of the $\delta$-potential is a very appealing one due to its relative simplicity and ease with which the necessary calculations can be carried out. It also reflects the essential properties of more complicated structures \cite{Belloni1}. The model is characterized by only one parameter, whose continuous variation from positive to negative values describes its repulsive or attractive strength; in particular, it possesses one localized level in the latter configuration. On the other hand, the quantum system, which at zero voltage is able to support bound orbitals, no longer has any discrete stationary levels when placed into the time and space unvarying electric field but instead only a continuum of states with their energies covering the whole axis, $-\infty<E<\infty$. As a result, the corresponding scattering matrix has poles only at the complex energies. A great deal of attention has been devoted to the analysis of the finite-width QW \cite{Moyer1,Bastard1,Austin1,Austin2,Austin3,Singh1,Borondo1,Ahn1,Austin4,Ghatak1,Ghatak2,Barrio1,Bloss1,Nakamura1,Juang1,Glasser2,Yuen1,delaCruz1,Kim1,Kuo1,Panda1,Zambrano1,Emmanouilidou1} and its $\delta$ counterpart \cite{Lukes1,Moyer2,Geltman1,Popov1,Scheffler1,Arrighini1,Fernandez1,Dargys1,Kundrotas1,Ludviksson1,Elberfeld1,Glasser1,Gottlieb1,Cocke1,Moyer3,Carpena1,Nickel1,Emmanouilidou2,Emmanouilidou3,Korsch1,Cavalcanti1,Alvarez1,Deych1,Galitski1,Moyer4,Brown1} subjected to a dc electric field with the outcome that one approach is sometimes completely at odds to the predictions of another scheme; in particular, the peculiarities of the transformation of the resonances into the true bound states at $\mathscr{E}\rightarrow0$ were scrutinized and the formation of new field-induced complex energy quasibound states and resonances was predicted.

Building on previous studies, below a detailed analysis was conducted of the resonances and quasibound states of the $\delta$-potential calculated from the three above-specified requirements. It can be seen that the complex energies derived as solutions of the stationary Schr\"{o}dinger equation with a non-zero field inevitably imply the exponential growth of the associated wave functions that, however, can be correctly construed with the help of the time-dependent picture. The development, based on the assumption that the energies have to remain real, leads to the conclusion that the infinitely small applied voltage at maximal scattering, $S=+1$, generates an infinite number of quasibound states in the positive energy continuum, the first set of which bears the features of electrons while the second exhibits the properties of a positively-charged particle that in solid state physics corresponds to hole excitations; for example, corpuscles residing in these two types of levels move in opposite directions as the field grows, and their evolution with increasing voltage ultimately forces them to coalesce with each other in what can be considered as the electric breakdown of the structure. The highest breakdown field is predicted to occur for mergers involving the level that developed from the zero-field bound state. Amalgamation of the levels is accompanied by the divergence of the associated dipole moments. This phenomenon, previously predicted only for the ground state \cite{Moyer3,Moyer4}, is calculated and analyzed in subsection \ref{QuasiboundStates1} for all quasibound levels. It should be noted that the corresponding eigenvalue equation has, at intensities $\mathscr{E}$ greater than the breakdown voltage, a pair of complex conjugate solutions that can be a mathematical indication of the formation in the electric field of a composite electron-hole-like structure. Corresponding maxima of the Wigner delay time were also computed as a function of the applied voltage. It can be seen that at the vanishing electric intensities, the predictions of the three methods coincide for the field-free bound state; however, even in this regime each approach has its own peculiarities. The difference between the results that coincided at $\mathscr{E}\ll1$ grows with the field. Contrary to the model of the $\delta$-potential that is symmetric at $\mathscr{E}=0$, the motion of the particle in the presence of the Robin wall takes place only on the half line. This lack of spatial symmetry has a drastic effect on the emergence and evolution of the quasibound states when the voltage is applied; in particular, for this system the field induces only hole-like quasibound levels and the lowest of them merges with the state that evolved from the field-free orbital that exists for the negative de Gennes distance, while the higher lying states survive any electric intensity.

\section{$\delta$-potential}\label{Sec_Delta}
The starting point of our analysis is the 1D stationary Schr\"{o}dinger equation
\begin{equation}\label{Schrodinger1}
\hat{H}\Psi(x)=E\Psi(x),
\end{equation}
where the Hamiltonian $\hat{H}$ is given by
\begin{equation}\label{Hamiltonian1}
\hat{H}=-\frac{\hbar^2}{2m}\frac{d^2}{dx^2}+V(x)-e\mathscr{E}x
\end{equation}
for the wave function $\Psi(x)$ of the particle with mass $m$ and charge $-e$ (with $e$ being an absolute value of the electronic charge) moving along an infinite straight line $-\infty<x\le+\infty$ in a uniform electric field $\mathscr{E}$ with a potential $V(x)$ being of the $\delta$-like form
\begin{equation}\label{DeltaPotential1}
V(x)=\frac{\hbar^2}{m}\frac{1}{\Lambda}\,\delta(x).
\end{equation}
Here, $\Lambda$ is a real coefficient , which has a dimension of length, being either positive or negative. Due to its presence, the matching conditions at $x=0$ are:
\begin{subequations}\label{MatchingCondtions1}
\begin{eqnarray}\label{MatchingCondtions1_1}
\Psi(0-)&=&\Psi(0+)\\
\label{MatchingCondtions1_2}
\Psi'(0+)-\Psi'(0-)&=&\frac{2}{\Lambda}\Psi(0)
\end{eqnarray}
\end{subequations}
with the prime denoting a derivative of the function with respect to its argument. Eq.~\eqref{MatchingCondtions1_2} demonstrates that there is a jump in the derivative of the wave function at the origin that is inversely proportional to the distance $\Lambda$. In the absence of an electric field, $\mathscr{E}=0$, the attractive potential, $\Lambda<0$, in addition to the continuous spectrum at $E>0$ (which is also characteristic for $\Lambda>0$) binds the particle at negative energy
\begin{equation}\label{EnergyDeltaZeroFields1}
E=-\frac{\hbar^2}{2m\Lambda^2},\quad\Lambda<0,
\end{equation}
while its normalized to unity,
\begin{equation}\label{Normalization1}
\int_{-\infty}^\infty\Psi^2(x)dx=1,
\end{equation}
wave function $\Psi(x)$ exponentially decreases away from the origin:
\begin{equation}\label{FunctionDeltaZeroFields1}
\Psi(x)=\frac{1}{|\Lambda|^{1/2}}\exp\!\left(-\left|\frac{x}{\Lambda}\right|\right).
\end{equation}

An applied electric field changes the charge distribution in the system. The quantitative measure of this influence is provided by the polarization, or dipole moment, $P(\mathscr{E})$, defined as \cite{Nguyen1,Olendski1,Olendski2}
\begin{equation}\label{Polarization1}
P(\mathscr{E})=\left\langle ex\right\rangle_\mathscr{E}-\left\langle ex\right\rangle_{\mathscr{E}=0},
\end{equation}
where the angular brackets denote a quantum mechanical expectation value:
\begin{equation}\label{AngleBrackets1}
\left\langle x\right\rangle=\int x\Psi^2(x)dx
\end{equation}
with the integration carried out over all available space, which in the case of the $\delta$-potential, reduces the polarization to
\begin{equation}\label{Polarization2}
P^\delta(\mathscr{E})=e\int_{-\infty}^\infty x\Psi^2(x)dx.
\end{equation}
As will be shown below, it is possible to calculate this quantity even in the case of open structures, such as the ones considered in the present study.

It is convenient to switch to dimensionless scaling from the outset so that all distances are measured in units of $|\Lambda|$, energies -- in units of $\hbar^2/\left(2m|\Lambda|^2\right)$, time -- in units of $2m|\Lambda|^2/\hbar$, polarization -- in units of $e|\Lambda|$, velocity -- in units of $\hbar/(2m|\Lambda|)$, electric fields -- in units of $\hbar^2/\left(2em|\Lambda|^3\right)$, and current density -- in units of $-e\hbar/\left(m|\Lambda|^2\right)$. Then, Eq.~\eqref{Schrodinger1} using the potential from Eq.~\eqref{DeltaPotential1} takes a universal form:
\begin{equation}\label{Schrodinger2}
-\Psi''(x)\pm\delta(x)\Psi(x)-\mathscr{E}x\Psi(x)=E\Psi(x),\\
\end{equation}
while the matching condition from Eq.~\eqref{MatchingCondtions1_2} is transformed to
\begin{equation}\label{MatchingCondtions2}
\Psi'(0+)-\Psi'(0-)=\pm2\Psi(0),
\end{equation}
where the upper (lower) sign refers to the repulsive (attractive) potential. The same convention will be used, as necessary, throughout the whole section.

Due to its generic definition, the scattering matrix describes the results of wave reflection from the structure when the total function $\Psi_t$ includes both the incoming [first term on the right-hand side of Eq.~\eqref{ScatteringFunction1}] and reflected (second item) components:
\begin{eqnarray}
\Psi_t(\mathscr{E};x)&=&{\rm Ci}^-\!\!\left(\!-\mathscr{E}^{1/3}x-\frac{E}{\mathscr{E}^{2/3}}\!\right)\nonumber\\
\label{ScatteringFunction1}
&+&S{\rm Ci}^+\!\!\left(\!-\mathscr{E}^{1/3}x-\frac{E}{\mathscr{E}^{2/3}}\!\right),\quad x\geq0.
\end{eqnarray}
Here, $${\rm Ci}^\pm(\eta)={\rm Bi}(\eta)\pm i{\rm Ai}(\eta)$$ (obviously, the superscript at ${\rm Ci}$ refers to the sign of its imaginary part and not to the $\delta$-potential), and ${\rm Ai}(\eta)$ and ${\rm Bi}(\eta)$ are Airy functions \cite{Abramowitz1,Vallee1}. To the left of the well, the fading at the negative infinity solution is:
\begin{equation}\label{WaveFunction3}
\Psi_n(x)=A_n{\rm Ai}\!\left(-\mathscr{E}^{1/3}x-\frac{E_n}{\mathscr{E}^{2/3}}\right),\quad x\leq0,
\end{equation}
where the explicitly-included subscript $n=0,1,2,\ldots$, counts the corresponding resonances (see below) and $A_n$ is a normalization constant. Matching according to Eqs.~\eqref{MatchingCondtions1_1} and \eqref{MatchingCondtions2} leads to the scattering matrix:
\begin{eqnarray}
&&S^{\delta\pm}(\mathscr{E};E)=\nonumber\\
\label{DeltaPotentialScatteringMatrix1}
&&-\frac{2{\rm Ai}\!\left(-\frac{E}{\mathscr{E}^{2/3}}\right){\rm Bi}\!\left(-\frac{E}{\mathscr{E}^{2/3}}\right)\pm\frac{\mathscr{E}^{1/3}}{\pi}-i\,2{\rm Ai}^2\!\left(-\frac{E}{\mathscr{E}^{2/3}}\right)}{2{\rm Ai}\!\left(-\frac{E}{\mathscr{E}^{2/3}}\right){\rm Bi}\!\left(-\frac{E}{\mathscr{E}^{2/3}}\right)\pm\frac{\mathscr{E}^{1/3}}{\pi}+i\,2{\rm Ai}^2\!\left(-\frac{E}{\mathscr{E}^{2/3}}\right)}.
\end{eqnarray}
Note that for real energies, this complex function, which also depends on the parameter $\mathscr{E}$, is unitary.

\subsection{Poles of the Scattering Matrix: Gamow-Siegert States}\label{GamowSiegertStates1}
Zeroing the denominator of the right-hand side of Eq.~\eqref{DeltaPotentialScatteringMatrix1} produces a universal equation for calculating the complex energies $E_{res_n}$:
\begin{equation}\label{DeltaPotentialEigenEquation1}
2\pi{\rm Ai}\!\left(\!-\frac{E_{res_n}}{\mathscr{E}^{2/3}}\!\right)\left[{\rm Bi}\!\left(\!-\frac{E_{res_n}}{\mathscr{E}^{2/3}}\!\right)+i{\rm Ai}\!\left(\!-\frac{E_{res_n}}{\mathscr{E}^{2/3}}\!\right)\right]\pm\mathscr{E}^{1/3}=0.
\end{equation}
Note that this equation can be derived in an alternative way; because the applied field is created by the potential that unrestrictedly decreases with $x\rightarrow+\infty$, it follows that at the non-zero voltage even for the attractive well there are, strictly speaking, no true bound states since the electron localized near the origin at $\mathscr{E}=0$ lowers its potential energy by tunneling away from the attractive center when the electric intensity is not zero. It is then contended \cite{Ahn1,delaCruz1,Kim1,Emmanouilidou1,Deych1,Galitski1,Yuen1} that the solution of the Schr\"{o}dinger equation should represent the outgoing waves at infinity and, since this requirement infers the non-zero imaginary component of the wave function $\Psi$, the energy also becomes complex. This results in a transformation of the true bound level into the resonance state with a finite lifetime
\begin{equation}\label{lifetime1}
\tau=\frac{1}{\Gamma},
\end{equation}
where the positive $\Gamma$ is a half width of the corresponding resonance:
\begin{equation}\label{ComplexEnergy1}
E_{res_n}=E_{r_n}-i\frac{\Gamma_n}{2}
\end{equation}
with $E_{r_n}$ being real. Accordingly, the function $\Psi_n(x)$ for positive $x$ is written as
\begin{equation}\label{WaveFunction2}
\Psi_n(x\geq0)=C_n{\rm Ci}^+\!\!\left(-\mathscr{E}^{1/3}x-\frac{E_{res_n}}{\mathscr{E}^{2/3}}\right),
\end{equation}
where $C_n$ is a normalization constant, and its match with the waveform from Eq.~\eqref{WaveFunction3} leads to Eq.~\eqref{DeltaPotentialEigenEquation1}.

From the properties of the Airy functions \cite{Abramowitz1,Vallee1} it is easy to derive the evolution of the field-free bound level at small electric intensities:
\begin{equation}\label{DeltaPotentialAsymptotics1}
{E_{res}^{\delta-}}_0(\mathscr{E})=-1-\frac{5}{16}\mathscr{E}^2-i\exp\!\left(\!-\frac{4}{3}\frac{1}{\mathscr{E}}\right),\quad\mathscr{E}\ll1.
\end{equation}
It can be seen that the real part of the energy depends quadratically on the voltage, which is a result of the symmetry of the field-free structure with respect to the inversion $x\rightarrow-x$.  An exponentially small increase of the half width
\begin{equation}\label{DeltaHalfWidthAsymptotics1}
{\Gamma_{res}^{\delta-}}_0=2\exp\!\left(\!-\frac{4}{3}\frac{1}{\mathscr{E}}\right),\quad\mathscr{E}\ll1,
\end{equation}
which is typical for a wide range of potentials that decay at infinity \cite{Galitski1}, physically means quite a small probability of tunneling away from the well and a particularly long lifetime of the state, as follows from Eq.~\eqref{lifetime1}.

\begin{figure}
\centering
\includegraphics[width=\columnwidth]{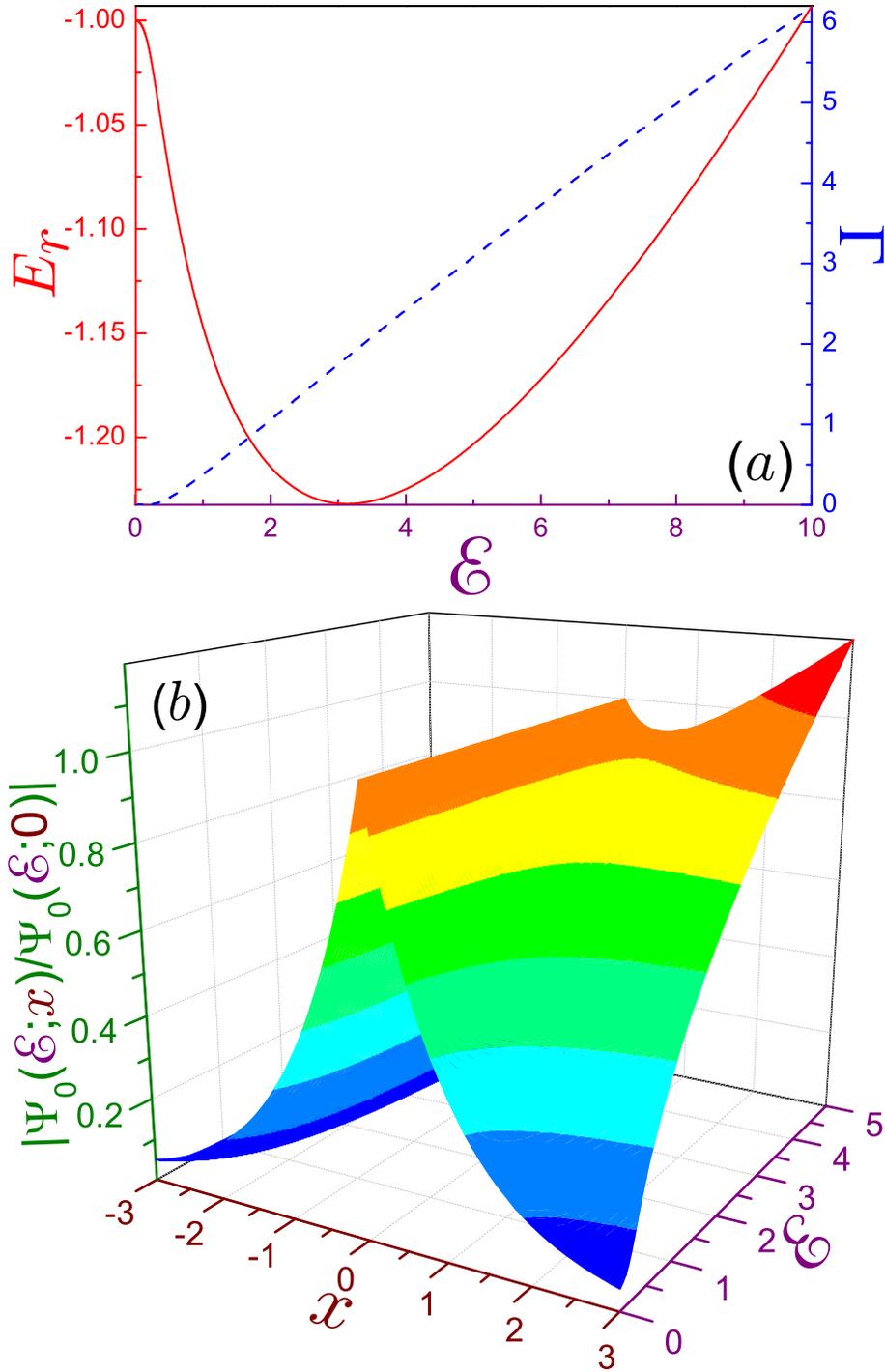}
\caption{\label{DeltaPotentialComplexEnergyFunction0}
(a) Real ${E_r}_0$ (solid line, left axis) and negative double imaginary $\Gamma_0$ (dashed curve, right axis) components of the complex energy as a function of the field calculated from Eq.~\eqref{DeltaPotentialEigenEquation1}. (b) Wave function $\Psi_0$ (normalized to its value at $x=0$) in terms of the distance $x$ and electric field $\mathscr{E}$.
}
\end{figure}

Panel (a) of Fig.~\ref{DeltaPotentialComplexEnergyFunction0} shows the zeroth level energy dependence on the field. A quadratic decrease of the real part and an exponentially small increase of the half width at the small intensities derived above are clearly seen in the plot. The real part of the energy reaches a minimum of ${E_r}_{min}=-1.232$ at $\mathscr{E}_{min}=3.125$, after which it demonstrates a permanent growth; in particular, it crosses its zero-field value of $-1$ at $\mathscr{E}=9.864$ and enters the positive part of the spectrum at $\mathscr{E}=26.303$. The half width $\Gamma_0$ after an almost flat profile at small $\mathscr{E}$ exhibits a nearly linear dependence on the applied field, which at higher voltages approaches $\mathscr{E}^{2/3}$. This growth implies a drastic decrease of the lifetime by field-enhanced tunneling.

The evolution of the associated wave function $\Psi_0(\mathscr{E};x)$ is depicted in panel (b) of Fig.~\ref{DeltaPotentialComplexEnergyFunction0}. Its most striking feature is an exponential growth with positive distance $x$ at the non-zero fields. Indeed, the claim that the function from Eq.~\eqref{WaveFunction2} describes the outgoing wave \cite{Ahn1,delaCruz1} is correct only for the {\em real} energies, but for the {\em complex} $E$ it inevitably leads to an exponential increase of the function at large $x$, which can be easily shown by employing the asymptotic behavior of the Airy functions \cite{Abramowitz1,Vallee1}. Complex eigenvalues were first introduced by Thomson \cite{Thomson1} in his analysis of the modes supported by an electric sphere and were later implemented in Gamow's explanation of alpha decay \cite{Gamow1}. The most serious criticism of these Gamow, or Siegert \cite{Siegert1}, states denies their physical legitimacy due to the unrestricted growth of the wave function, with the vivid manifestation of this 'exponential catastrophe' presented in Fig.~\ref{DeltaPotentialComplexEnergyFunction0}. However, this divergence is elegantly eliminated by considering the temporal evolution of the structure; as a matter of fact, if the time decay of the state is considered and even the corresponding lifetime from Eq.~\eqref{lifetime1} is introduced, it is logical to investigate the associated time-dependent function ${\bm \Psi}(\mathscr{E};x,t)=\Psi(\mathscr{E};x)e^{-iE_{res}t}$. For the large negative argument in the waveform from Eq.~\eqref{WaveFunction2}, the total function reads:
\begin{eqnarray}
{\bm \Psi}_0(\mathscr{E};x,t)&=&C_0\frac{\mathscr{E}^{1/6}}{\pi^{1/2}\left(E_{r_0}+\mathscr{E}x-i\frac{\Gamma_0}{2}\right)^{1/4}}\nonumber\\
&\times&\exp\!\left(i\left[\frac{2}{3}\frac{(E_{r_0}+\mathscr{E}x)^{3/2}}{\mathscr{E}}-E_{r_0}t+\frac{\pi}{4}\right]\right)\nonumber\\
\label{WaveFunction4}
&\times&\exp\!\left(-\frac{\Gamma_0}{2}\left[t-T(\mathscr{E};x)\right]\right),\quad\mathscr{E}x\gg1,
\end{eqnarray}
where also a smallness of the half width $\Gamma_0$ has been assumed. The first exponent in Eq.~\eqref{WaveFunction4} is a plane wave that describes free motion in the linear potential and $T(\mathscr{E};x)$ is just the classical time needed for the electron to travel the distance between the quasi classical turning point $x_{qc}=-E_{r_0}/\mathscr{E}$ and the coordinate $x>x_{qc}$:
\begin{subequations}\label{Classical1}
\begin{align}\label{Classical1_Time1}
T(\mathscr{E};x)&=\frac{(E_{r_0}+\mathscr{E}x)^{1/2}}{\mathscr{E}}.
\intertext{This is particularly clear when this expression is rewritten in the alternative equivalent form:}
\label{Classical1_Time2}
T(x)&=2\,\frac{x-x_{qc}}{v(x)},
\end{align}
\end{subequations}
where $v(x)\equiv v(\mathscr{E};x)$ is a field-dependent classical speed at $x$:
\begin{equation}\label{ClassicalVelocity1}
v(\mathscr{E};x)=2\left(E_{r_0}+\mathscr{E}x\right)^{1/2}.
\end{equation}
The above equations in this paragraph, similar to the analysis of alpha decay \cite{Bohm1,Baz1,Holstein1}, can be construed as follows. In the derivation of Eq.~\eqref{WaveFunction4} it was tacitly assumed that it is valid for all times $-\infty<t<\infty$. However, in reality the decay does not start in the infinitely remote past since the corresponding state has to be created first by, say, the adiabatic varying of the field or any other means. Accordingly, it is natural to choose as the origin the moment when the emitted particle emerges at the turning point $x_{qc}$ after tunneling through the triangular barrier, which the electron with the negative energy $E_{r_0}<0$ located inside the $\delta$-well "sees" to its right. This also means that at this point, the prehistory of the formation of the scattering level at $t<0$ is of no concern. Then, at any positive time the particle travels with an average speed $\overline{v}=v(x)/2$ to reach the observation point $x$ at moment $T(x)$ from Eq.~\eqref{Classical1_Time2}. Consequently, it does make sense to talk about measuring the probability density $\rho(\mathscr{E};x,t)\equiv\left|{\bm \Psi}(\mathscr{E};x,t)\right|^2$ at detector position $x$ at times $t\ge T(x)$ only:
\begin{eqnarray}
&&\rho_0(\mathscr{E};x,t)=\nonumber\\
\label{Density1}
&&\frac{|C_0|^2\mathscr{E}^{1/3}}{\pi\!\left[\left(E_{r_0}\!+\!\mathscr{E}x\right)^2\!\!\!+\!\left(\frac{\Gamma_0}{2}\right)^2\right]^{1/4}}\,e^{-\Gamma_0\left[t-T(\mathscr{E};x)\right]},\, t\ge T(\mathscr{E};x).
\end{eqnarray}
The corresponding current density in the $x$ direction
\begin{equation}\label{CurrentDensity2}
j_x=\!\!\!\left[v(x)-\frac{1}{8}\frac{\mathscr{E}\Gamma_0}{\left(E_{r_0}+\mathscr{E}x\right)^2+\left(\frac{\Gamma_0}{2}\right)^2}\right]\!\!\rho_0(\mathscr{E};x,t),
\end{equation}
which is calculated from the general expression \cite{Landau1}
\begin{equation}\label{CurrentDensity1}
{\bf j}={\rm Im}({\bm\Psi}^\ast{\bm\nabla}{\bm\Psi}),
\end{equation}
apart from the familiar velocity-dependent term [first item in Eq.~\eqref{CurrentDensity2}], contains an additional contribution  that is proportional to the half width $\Gamma$. Such a viewpoint eliminates the 'exponential catastrophe', yielding instead the anticipated  exponential decay law \cite{Bohm1,Baz1,Holstein1} with its lifetime taken from Eq.~\eqref{lifetime1}. However, neglecting all times smaller than $T(x)$ and, in particular, $t<0$, completely ignores the processes of under-barrier tunneling at $0<x<x_{qc}$ and this semi classical reasoning is therefore not a strictly quantum one. To correctly account for the build-up of the levels at earlier times $t<0$, a solution of Eq.~\eqref{DeltaPotentialEigenEquation1} with the positive imaginary component, which is a complex conjugate of its counterpart from Eq.~\eqref{ComplexEnergy1}, is required. Negative half widths $\Gamma$ specify the system in the growing state, which is called antiresonance (see Sec. \ref{QuasiboundStates1}). They are also appropriate to describe a particle traveling back in time towards the past. A separation of the complex energy eigenfunctions into those corresponding to the physical states at the earlier times (with $\Gamma<0$) and those associated with the configuration for the later times, $\Gamma>0$, is a peculiar property of the Gamow vectors with complex energies \cite{Bohm2}. From the point of view of mathematical formalism, the Gamow-Siegert states  do not represent vectors from the Hilbert space of the quantum structure under consideration, being instead eigenvectors of the rigged Hilbert space (see Refs. \cite{Bohm2,Civitarese1,deLaMadrid1} and references therein). 

In addition to the resonance that, at vanishing electric intensities, transforms into the field-free bound state, Eq.~\eqref{DeltaPotentialEigenEquation1} has other solutions for either attractive \cite{Ludviksson1,Alvarez1} or repulsive potentials. The easiest way to show their existence is to substitute into the equation an Airy functions relation \cite{Abramowitz1,Vallee1}
$$
{\rm Ai}(z)\mp i{\rm Bi}(z)=2e^{\mp i\pi/3}{\rm Ai}\!\left(ze^{\pm i2\pi/3}\right)
$$
and to consider the resulting transcendental formula
\begin{equation}\label{DeltaPotentialEigenEquation3}
4\pi ie^{-i\pi/3}{\rm Ai}\!\left(-\frac{E}{\mathscr{E}^{2/3}}\right){\rm Ai}\!\left(-\frac{E}{\mathscr{E}^{2/3}}\,e^{i2\pi/3}\right)\pm\mathscr{E}^{1/3}=0
\end{equation}
in the limit of the low voltages. After some algebra involving the properties of the Airy functions, two sets of solutions are achieved, which at $\mathscr{E}\ll1$ are:
\begin{subequations}\label{DeltaComplexSolutionsTwoSets1}
\begin{eqnarray}\label{DeltaComplexSolutionsTwoSets1_Set1}
E_{res_n}^{(1)\pm}&=&-a_n\mathscr{E}^{2/3}\mp\frac{1}{2}\,\mathscr{E}\!+\!\frac{1}{4}\frac{{\rm Bi}'(a_n)}{{\rm Bi}(a_n)}\mathscr{E}^{4/3}\!-\!i\frac{\mathscr{E}^{4/3}}{4\pi{\rm Bi}^2(a_n)}\\
\label{DeltaComplexSolutionsTwoSets1_Set2}
{E_{res_n}^{(2)\pm}}&=&\frac{1}{2}\,a_n\mathscr{E}^{2/3}\pm\frac{1}{2}\,\mathscr{E}+i\,\frac{3^{1/2}}{2}\,a_n\mathscr{E}^{2/3}.
\end{eqnarray}
\end{subequations}
Here, (all negative) coefficients $a_n$, $n=1,2,\ldots$, are solutions of equation ${\rm Ai}(a_n)=0$ \cite{Abramowitz1,Vallee1}. Note the opposite signs of the real parts of the energies $E_{res_n}^{(1)\pm}$ and $E_{res_n}^{(2)\pm}$ and different powers of the field dependence of their imaginary components. We will address more of the properties of these states while comparing them to the results obtained via other methods.

\subsection{Real-energy Quasibound States: $S=+1$}\label{QuasiboundStates1}
Having seen the properties of the complex Gamow-Siegert  states, let us now return to the scattering matrix from Eq.~\eqref{DeltaPotentialScatteringMatrix1}. To remain rigorously within the time-independent quantum treatment without divergences, only the {\em real} energies $E$ will be used in this and the following subsections, unless otherwise stipulated. For our geometry, the matrix $S$ characterizes the influence of the $\delta$-potential on the incident particle; namely, it is well known \cite{Landau1} that without it, $\Lambda=0$, the corresponding waveform is proportional to the Airy function ${\rm Ai}(\eta)$, which corresponds to $S=-1$ in Eq.~\eqref{ScatteringFunction1}:
\begin{equation}\label{ScatteringFunction0}
\Psi_{\Lambda=0}(\mathscr{E};x)=-2i{\rm Ai}\!\left(\!-\mathscr{E}^{1/3}x-\frac{E}{\mathscr{E}^{2/3}}\!\right).
\end{equation}
Then, in the superposition of both forces, the scattered wave $\Psi_{sc}$ is the difference between the total function from Eq.~\eqref{ScatteringFunction1} and its unperturbed counterpart, Eq.~\eqref{ScatteringFunction0}:
\begin{eqnarray}
\Psi_{sc}(\mathscr{E};x)&=&\Psi_t(\mathscr{E};x)-\Psi_{\Lambda=0}(\mathscr{E};x)\nonumber\\
\label{ScatteringFunction2}
&=&(1+S){\rm Ci}^+\!\!\left(\!-\mathscr{E}^{1/3}x-\frac{E}{\mathscr{E}^{2/3}}\!\right).
\end{eqnarray}
This equation shows that $\Psi_{sc}$ is a purely outgoing wave. The squared modulus of its amplitude with its maximum normalized to unity is called the scattering probability:
\begin{equation}\label{ScatProbab1}
p(\mathscr{E};E)=\frac{1}{4}\,\left|1+S(\mathscr{E};E)\right|^2.
\end{equation}
Note that the extremely localized potential, as expected, does not scatter the wave of the arbitrary strength electric field when its position coincides with one of the nodes of the unperturbed orbital $\Psi_{\Lambda=0}$:
\begin{equation}\label{ZeroScattering1}
p^\delta\!\left(\mathscr{E};-a_n\mathscr{E}^{2/3}\right)=0.
\end{equation}
In this way, the significance of the special value $-1$ of the matrix $S$ mentioned in the Introduction is established. Another important property is the identical vanishing of the current density of the total function $\Psi_t$ at any energy, as it easily follows from its substitution into Eq.~\eqref{CurrentDensity1}. Thus, mathematical models of the Gamow-Siegert states and that based on the real energy analysis describe different physical situations: while the former one calculates the temporal leakage from the well of the level that was prepared at the earlier times, the subject of the latter method is the stationary configuration that emerged as a result of the interference between the incident and reflected waves, with the resulting net current being an exact zero. As a consequence of this, {\em real} values of the energy guarantee that the waveform $\Psi_t(x)$ is finite everywhere.

Distortion of the electron motion by the $\delta$-potential is maximal for energies $E_n$, $n=0,1,2,\ldots$, at which the  scattering matrix $S^\delta$ changes to positive unity:
\begin{equation}\label{DeltaPotentialEigenEquation2}
2\pi{\rm Ai}\!\left(\!-\frac{E}{\mathscr{E}^{2/3}}\!\right){\rm Bi}\!\left(\!-\frac{E}{\mathscr{E}^{2/3}}\!\right)\pm\mathscr{E}^{1/3}=0,
\end{equation}
and the associated total function $\Psi_t(x)$ to the right degenerates to the Airy function ${\rm Bi}(\eta)$:
\begin{equation}\label{DeltaFunction0}
\Psi_{t_n}(x)=B_n{\rm Bi}\!\left(-\mathscr{E}^{1/3}x-\frac{E_n}{\mathscr{E}^{2/3}}\right),\quad x\geq0.
\end{equation}
For the attractive potential, Eq.~\eqref{DeltaPotentialEigenEquation2} was derived previously with the help of the Green functions \cite{Glasser1,Carpena1} without any detailed analysis. The same configuration was discussed by C. A. Moyer \cite{Moyer3,Moyer4}, who concentrated on finding the energy and associated polarization of its lowest level only, which at the vanishing electric intensities tends to the zero-field bound state:
\begin{subequations}\label{DeltaEnergiesSmallFields1}
\begin{align}\label{DeltaEnergiesSmallFields1B0}
E_0&=-1-\frac{5}{16}\,\mathscr{E}^2,\quad \mathscr{E}\ll1.\\
\intertext{In addition, for either the $\delta$-well or barrier, Eq.~\eqref{DeltaPotentialEigenEquation2} has two infinite sets of positive solutions that will be denoted below by the superscripts $A$ or $B$ corresponding to their behavior at low voltages; namely, for the weak fields, $\mathscr{E}\ll1$, one finds:}
\label{DeltaEnergiesSmallFields1An}
E_n^{A\pm}&=-a_n\mathscr{E}^{2/3}\mp\frac{1}{2}\,\mathscr{E}+\frac{1}{4}\frac{{\rm Bi}'(a_n)}{{\rm Bi}(a_n)}\mathscr{E}^{4/3},\\
\label{DeltaEnergiesSmallFields1Bn}
E_n^{B\pm}&=-b_n\mathscr{E}^{2/3}\pm\frac{1}{2}\,\mathscr{E}+\frac{1}{4}\frac{{\rm Ai}'(b_n)}{{\rm Ai}(b_n)}\mathscr{E}^{4/3},
\end{align}
\end{subequations}
$n=1,2,\ldots$. Here, negative $b_n$ is the $n$th zero of the Airy function ${\rm Bi}(x)$: ${\rm Bi}(b_n)=0$ \cite{Abramowitz1,Vallee1}. It is important to stress that Eq.~\eqref{DeltaPotentialEigenEquation2}, in addition to the real solutions, has complex roots too. This follows from the fact that at low electric intensities the $B$ set is essentially determined by ${\rm Bi}(\eta_0)=0$ with $\eta_0=-E/\mathscr{E}^{2/3}$. This equation is also satisfied by the complex numbers $\beta_n$, in addition to the real coefficients $b_n$ \cite{Abramowitz1,Vallee1}. However, these states are disregarded due to the convention of avoiding the 'exponential catastrophe'. Therefore, under the assumption of keeping the energies {\em real} we have found that quantization results in a countably infinite number of solutions.  To place them correctly within the nomenclature of the other solutions of the Schr\"{o}dinger equation, it should be noted that historically, the terms "quasibound (or quasi-stationary) state" and "resonance" were used interchangeably to describe the {\em complex}-energy Gamow-Siegert level. The standard procedure for categorizing the poles of the $S$-matrix implements their location in the complex $k$-plane, where $k\equiv k_r+ik_i=\sqrt{E}$ with real $k_r$ and $k_i$ \cite{Kukulin1,Bang1}. Bound states, $E<0$, lie on the imaginary semi axis in the upper $k$-halfplane, $k_{r_B}=0$, $k_{i_B}=+\sqrt{|E|}>0$, which leads to fading  function at large distances, $\Psi_B(x)\xrightarrow[|x|\rightarrow\infty]{}\exp\left(-k_{i_B}|x|\right)$, as expected. A complete mathematical set of solutions should also include those dependencies with the purely imaginary {\em negative} wave vector, $k_{r_{AB}}=0$, $k_{i_{AB}}=-\sqrt{|E|}<0$. Accordingly, the waveforms $\Psi_{AB}$ of such {\em anti}-bound states \cite{Kukulin1,Bang1} diverge at infinity: $\Psi_{AB}(x)\xrightarrow[x\rightarrow\infty]{}\exp\left(\left|k_{i_{AB}}\right|x\right)$. Despite this unlimited growth, these levels can have a physical meaning too \cite{Ohanian1,Heiss1}. All other poles of the scattering matrix lie in the lower $k$-halfplane, $k_i<0$, with the positive real part of the wave vector corresponding to the above-discussed Gamow quasi-stationary states (resonances) while those with $k_r<0$ are called antiresonances and describe an ingoing wave \cite{Kukulin1}. Obviously, solutions from Eqs.~\eqref{DeltaEnergiesSmallFields1} do not fall into one of these categories, since in the first instance they are not poles of the $S$-matrix and, second, all their energies (except the lowest one of the attractive potential) are positive. Moreover, the corresponding functions at large positive $x$ neither exponentially diverge nor fade, presenting instead oscillatory damped modes, as it follows from Eq.~\eqref{DeltaFunction0} and asymptote of the Airy function. Nevertheless, in the recent analysis of the lowest level evolution in the field \cite{Moyer4} this orbital was called the quasibound state and was defined broadly as the level having a connectedness to the true bound state through the variation of some physical parameter. In our extension to all the solutions (including those with $E\geq0$), we found it relevant to call them real energy quasibound (REQB) states. These are bound states, since for each of them the wave function $\Psi_n(x)$ has a finite absolute value everywhere including the point $x=+\infty$ while the prefix 'quasi' in their definition means that due to the slow decrease at large positive $x$ of the function ${\rm Bi}$ from Eq.~\eqref{DeltaFunction0}, they cannot be normalized according to Eq.~\eqref{Normalization1} \cite{Moyer4}. Contrary to these states, the divergent-at-infinity Gamow-Siegert solutions will be called resonances \cite{Kukulin1,Moyer4}. It is important to underline that the REQB states with the discrete energies $E_0$, $E_n^A$, and $E_n^B$ are embedded into the continuum of the delocalized levels, with its energies ranging from the negative to positive infinity; as such, they mathematically represent only a measure zero part of all continuum energy eigenstates of the given Hamiltonian, while physically they describe  the largest possible, $p=S=1$, disturbance by the $\delta$-potential of the motion in the uniform electric field. To end this part of the discussion, it should be mentioned that, in addition to the levels discussed above, open systems can also support under very special conditions true bound states in the continuum, i.e., waves that remain localized (with square integrable functions) even though they coexist with a continuous spectrum of radiating oscillations that can carry energy away \cite{Hsu1}.

The first term on the right-hand side of Eq.~\eqref{DeltaEnergiesSmallFields1An} states that the $A$ set of solutions reflects the creation by the applied voltage and the $\delta$-potential of the triangular QW \cite{Olendski2,Olendski1,Katriel1}, with the wave function taken from Eq.~\eqref{WaveFunction3} and the subsequent terms representing an admixture due to its coupling to the right half space. Since the leading term in this formula is independent of the sign of the $\delta$ term, the formation of the triangular well takes place at either a positive or negative electrostatic potential $V(x)$ while the interaction between the left and right semi-infinite areas carries its sign. In the same way, the first expression on the right-hand side of Eq.~\eqref{DeltaEnergiesSmallFields1Bn} describes the formation in the region $x\ge0$ terminated by the impenetrable wall of the standing wave from Eq.~\eqref{DeltaFunction0} with the linear in the field and higher order factors there describing the correction due to coupling to the left-hand territory. The first part of the mathematical inequality chain 
\begin{equation}\label{InequalityChain1}
|a_{n+1}|>|b_{n+1}|>|a_n|
\end{equation}
physically means that the $A$ solutions are located to the left of the $\delta$-potential, with the energies $E_n^A$ being larger than those of the corresponding $B$ states with their distribution spreading at $x>0$. The sequence from Eq.~\eqref{InequalityChain1} also states that the $B$ and $A$ levels alternate on the energy axis.

The product of the two Airy functions on the left-hand side of Eq.~\eqref{DeltaPotentialEigenEquation2} is a bounded function of the energy: it decreases to zero as $\mathscr{E}^{1/3}/|E|^{1/2}$ for large negative $E$, reaches its global maximum at $E=0$, and for positive energies it presents sinusoidal oscillations with its amplitude modulated again by the same factor $\mathscr{E}^{1/3}/E^{1/2}$. Hence, for quite large electric intensities, this equation does not have any solutions. The disappearance of the levels at increasing voltage occurs as a coalescence of the two adjacent states at the electric fields $\mathscr{E}_n^\times$ that are different for the well and the barrier:
\begin{subequations}\label{DeltaCoalesceField1}
\begin{eqnarray}\label{DeltaCoalesceFieldMinus1}
\mathscr{E}_n^{\times-}&=&8\pi^3f^3(s_{2n})\\
\label{DeltaCoalesceFieldPlus1}
\mathscr{E}_n^{\times+}&=&-8\pi^3f^3(s_{2n+1})
\end{eqnarray}
\end{subequations}
and the energies $E_n^{\times\pm}$ at the merger are:
\begin{subequations}\label{DeltaCoalesceEnergy1}
\begin{eqnarray}\label{DeltaCoalesceEnergyMinus1}
E_n^{\times-}&=&-s_{2n}\left(\mathscr{E}_n^{\times-}\right)^{2/3}=-4\pi^2s_{2n}f^2(s_{2n})\\
\label{DeltaCoalesceEnergyPlus1}
E_n^{\times+}&=&-s_{2n+1}\left(\mathscr{E}_n^{\times+}\right)^{2/3}=-4\pi^2s_{2n+1}f^2(s_{2n+1}),
\end{eqnarray}
\end{subequations}
where 
\begin{equation}\label{function_f1}
f(s)={\rm Ai}(s){\rm Bi}(s)
\end{equation}
and non-positive $s_n$ is the $n$th solution, $n=0,1,2,\ldots$, of equation $f'(s)=0$, or in the expanded form
\begin{equation}\label{DeltaCrossingEquation1}
{\rm Ai}(s){\rm Bi}'(s)+{\rm Ai}'(s){\rm Bi}(s)=0.
\end{equation}
Since $s_0=0$ \cite{Abramowitz1,Vallee1}, the breakdown field of the two lowest levels is
\begin{equation}\label{DeltaCoalesceFieldMinus0}
\mathscr{E}_0^{\times-}=\mathscr{E}_f
\end{equation}
with
\begin{equation}\label{FundamentalField1}
\mathscr{E}_f=\frac{1}{3}\frac{\Gamma^3(1/3)}{\Gamma^3(2/3)}=2.58106\ldots,
\end{equation}
where $\Gamma(x)$ is the $\Gamma$-function \cite{Abramowitz1}. For quite large $n$ it is elementary to derive an asymptote
\begin{equation}\label{AsymptotSn1}
s_n=-\left(\frac{3}{4}\pi n\right)^{\!2/3},\quad n\gg1,
\end{equation}
that leads to the approximate formula  for $\mathscr{E}_n^\times$:
\begin{subequations}\label{DeltaCrossingField1}
\begin{eqnarray}\label{DeltaCrossingFieldMinus1}
\mathscr{E}_n^{\times-}&=&\frac{2}{3\pi}\frac{1}{n},\quad n\gg1\\
\label{DeltaCrossingFieldPlus1}
\mathscr{E}_n^{\times+}&=&\frac{4}{3\pi}\frac{1}{2n+1},\quad n\gg1.
\end{eqnarray}
\end{subequations}
Table~\ref{Table1} lists the exact values of $s_n$ and their approximations by Eq.~\eqref{AsymptotSn1} together with the exact and approximate coalescence fields and energies. It shows that the estimates from Eqs.~\eqref{AsymptotSn1} and ~\eqref{DeltaCrossingField1} provide reasonably good accuracy, even for small $n$.

\newpage
\begin{sidewaystable}
%
\caption{Exact and approximate solutions $s_n$ of Eq.~\eqref{DeltaCrossingEquation1} together with the dissociation fields $\mathscr{E}_n^\times$ and energies $E_n^\times$ for the $\delta$-potential}
\centering 
\begin{tabular}{c|c c|c c|c||c c|c c|c}
\hline
\hline
\multirow{2}[3]{*}{$n$}&\multicolumn{2}{c|}{$s_{2n}$}&\multicolumn{2}{c|}{$\mathscr{E}_n^{\times-}$}&\multirow{2}[3]{*}{$E_n^{\times-}$}&\multicolumn{2}{c|}{$s_{2n+1}$}&\multicolumn{2}{c|}{$\mathscr{E}_n^{\times+}$}&\multirow{2}[3]{*}{$E_n^{\times+}$}\\[0.5ex]
\cline{2-5}\cline{7-10}
&Exact&Eq.~\eqref{AsymptotSn1}&Exact&Eq.~\eqref{DeltaCrossingFieldMinus1}& &Exact&Eq.~\eqref{AsymptotSn1}&Exact&Eq.~\eqref{DeltaCrossingFieldPlus1}\\ 
\hline
0&0&0&2.58106&$\infty$&0&-1.76475&-1.77068&0.39591&0.42441&0.95150\\
1&-2.80824&-2.81078&0.20773&0.21221&0.98501&-3.68166&-3.68317&0.14006&0.14147&0.99293\\
2&-4.46080&-4.46184&0.10549&0.10610&0.99593&-5.1767&-5.1775&0.084566&0.084883&0.99736\\
3&-5.84606&-5.84667&0.070551&0.070736&0.99815&-6.47897&-6.47947&0.060514&0.060630&0.99864\\
4&-7.08232&-7.08273&0.052973&0.053052&0.99895&-7.66095&-7.66130&0.047102&0.047157&0.99917\\
5&-8.21847&-8.21878&0.042401&0.042441&0.99933&-8.75768&-8.75795&0.038553&0.038583&0.99944\\
6&-9.28076&-9.28100&0.035344&0.035368&0.99953&-9.78949&-9.78971&0.032629&0.032647&0.99960\\
7&-10.28532&-10.28552&0.030300&0.030315&0.99966&-10.76947&-10.76965&0.028282&0.028294&0.99970\\
8&-11.24297&-11.24313&0.026516&0.026526&0.99974&-11.70670&-11.70685&0.024957&0.024965&0.99977\\
9&-12.16141&-12.16155&0.023572&0.023579&0.99979&-12.60778&-12.60791&0.022332&0.022338&0.99981\\
10&-13.04638&-13.04650&0.021216&0.021221&0.99983&-13.47773&-13.47784&0.020206&0.020210&0.99985\\
20&-20.70998&-20.71003&0.010610&0.010610&0.99996&-21.05373&-21.05377&0.010351&0.010352&0.99996\\
50&-38.14819&-38.14820&0.0042441&0.0042441&0.99999&-38.40209&-38.40210&0.0042021&0.0042021&0.99999\\
100&-60.55649&-60.55650&0.0021221&0.0021221&1&-60.75818&-60.75819&0.0021115&0.0021115&1\\
[1ex]
\hline
\end{tabular}
\label{Table1}
\end{sidewaystable}
\begin{figure*}
\centering
\includegraphics[width=\columnwidth]{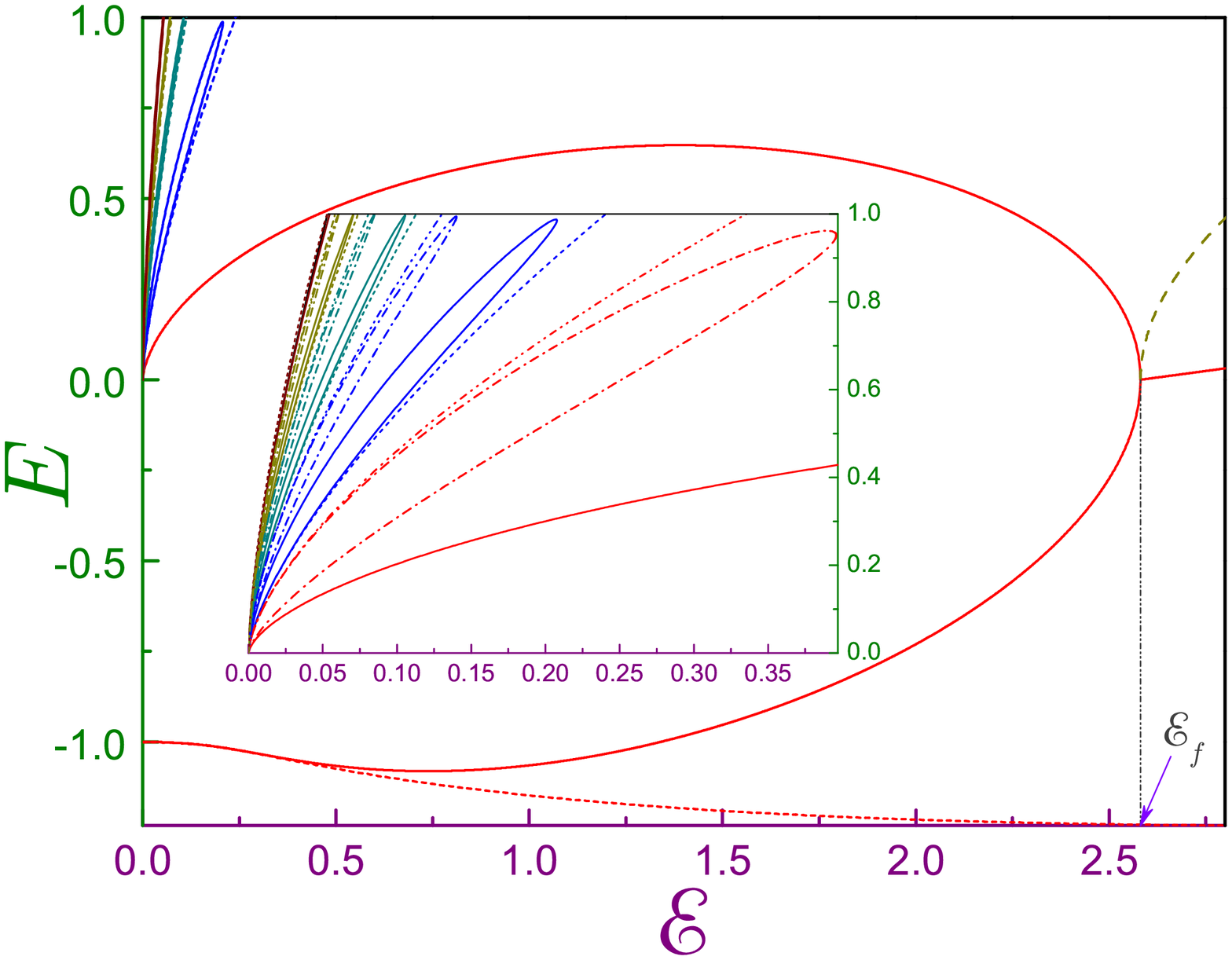}
\caption{\label{DeltaEnergies}
Energies $E$ of the quasibound states of the attractive $\delta$-potential (solid lines) as a function of the electric field $\mathscr{E}$. The upper edge of each petal corresponds to the $B$ level. Dotted curves depict real components of the complex solutions of Eq.~\eqref{DeltaPotentialEigenEquation1}. The thin vertical dash-dot-dotted line shows the location of the fundamental dissociation field $\mathscr{E}_f$. Solid and dashed lines at $\mathscr{E}>\mathscr{E}_f$ denote real and positive imaginary parts, respectively, of the first complex solution of Eq.~\eqref{DeltaPotentialEigenEquation2}. The inset depicts an enlarged view of several positive energy states, where levels of the repulsive barrier from Eq.~\eqref{DeltaPotentialEigenEquation2} (dash-dotted curves) and ~\eqref{DeltaPotentialEigenEquation1} (dash-dot-dotted lines) are also shown. Contrary to the attractive well, in the latter geometry the lower rim of each petal is formed by the $B$ level.}
\end{figure*}

Fig.~\ref{DeltaEnergies} shows the energies of REQB states calculated from Eq.~\eqref{DeltaPotentialEigenEquation2} together with the real parts of the complex energies being solutions of Eq.~\eqref{DeltaPotentialEigenEquation1}. At the weak fields, the energy $E_0$ decreases quadratically with electric intensity, as derived in Eq.~\eqref{DeltaEnergiesSmallFields1B0}. This can be interpreted as an increase in the binding of the electron by the small voltages. The ground state energy reaches a minimum of ${E_0}_{min}=-1.0806$ at ${\mathscr{E}_0}_{min}=0.739$. A comparison with the corresponding Gamow-Siegert data provided above shows a conspicuous difference at these fields, while for small $\mathscr{E}$ the energies calculated by either method are the same. The  deviation of the energies at $\mathscr{E}\gtrsim0.5$ is clearly seen in the figure. A subsequent increase of the voltage leads to the growth of the energy until it approaches zero at $\mathscr{E}_f$, where it amalgamates with the lowest $B$ level. All positive energies grow from their zero value as $\mathscr{E}^{2/3}$ at the weak fields, with the steepness being higher for larger $n$. This energy increase for the electric potential that seemingly has to force them downward is explained for the $A$ levels by the formation of the triangular QW, as discussed above. The lowest $B$ level that does not have its complex counterpart passes at ${\mathscr{E}_1}_{max}=1.372$ through the broad maximum of ${E_1}_{max}=0.6475$, after which it decreases towards zero. To determine the energy behavior close to this merger, it is convenient to represent Eq.~\eqref{DeltaPotentialEigenEquation2} in the parametric form \cite{Moyer3}
\begin{subequations}\label{Parametric1}
\begin{eqnarray}\label{ParametricEnergy1}
E&=&-4\pi^2zf^2(z)\\
\label{ParametricField1}
\mathscr{E}&=&8\pi^3f^3(z),
\end{eqnarray}
\end{subequations}
where the coefficient $z$, which is equal to $z=-E/\mathscr{E}^{2/3}$, varies from zero to positive infinity (for the lower level) or to $b_1$ (for the upper state). Close to the coalescence, this parameter is small, $|z|\ll1$, and the Taylor expansion simplifies Eqs.~\eqref{Parametric1} to
\begin{subequations}\label{Parametric2}
\begin{eqnarray}\label{ParametricEnergy2}
E&=&-4\pi^2f^2(0)z\\
\mathscr{E}&=&8\pi^3\left[f^3(0)+\frac{3}{2}f^2(0)f''(0)z^2\right]\nonumber\\
\label{ParametricField2}
&=&\left[1+\frac{3}{2}\frac{f''(0)}{f(0)}z^2\right]\mathscr{E}_f.
\end{eqnarray}
\end{subequations}
Eliminating $z$ from these equations, one gets after some simple algebra
\begin{equation}\label{AsymptotCriticalField1}
E_{\left\{^0_1\right\}}=\mp\left(\frac{\mathscr{E}_f}{3}\right)^{\!\!1/2}\left(\mathscr{E}_f-\mathscr{E}\right)^{1/2},\quad\mathscr{E}\rightarrow\mathscr{E}_f.
\end{equation}
For the higher lying amalgamations, this expression is generalized as
\begin{eqnarray}
E=E_n^\times\left[1\mp\frac{1}{s_n}\!\left(\!-\frac{2}{3}\frac{f(s_n)}{f''(s_n)}\frac{1}{\mathscr{E}_n^\times}\!\right)^{\!\!1/2}\!\!\left(\mathscr{E}_n^\times-\mathscr{E}\right)^{1/2}\right],\nonumber\\
\label{AsymptotCriticalField1_1}
\mathscr{E}\rightarrow\mathscr{E}_n^\times,\quad n\geq1,
\end{eqnarray}
which is also valid for the repulsive potential. The coalescence of the two states physically results in ionization of the structure by the growing field when the $\delta$-potential can no longer bind the charged particle. Higher lying states dissociate at weaker electric fields, as they are less bound by the potential. As Fig.~\ref{DeltaEnergies} demonstrates, for larger quantum numbers $n$ the energies of the $A$ levels deviate less from their complex-Airy-functions counterparts, while the petals formed by the $A$ and $B$ energies become narrower. For comparison, the curves of the repulsive potential are also shown in the inset. In this case, the energies can take positive values only with all basic features described for the QW being observed too. One major difference lies in the fact that for the barrier, the lower edge of each petal is formed by the $B$ level. Additionally, it should be noted that to the right of the field $\mathscr{E}_n^\times$ at which the levels with real energies merge,  Eq.~\eqref{DeltaPotentialEigenEquation2} has two complex conjugate solutions. Real and positive imaginary components of the lowest levels are also shown in the figure. The real part grows with the field from its value $E_n^\times$ at the breakdown while the magnitude of the imaginary part increases more rapidly from its zero value. The physical interpretation of these mathematically correct solutions of Eq.~\eqref{DeltaPotentialEigenEquation2} will be discussed below.

\begin{figure}
\centering
\includegraphics[width=\columnwidth]{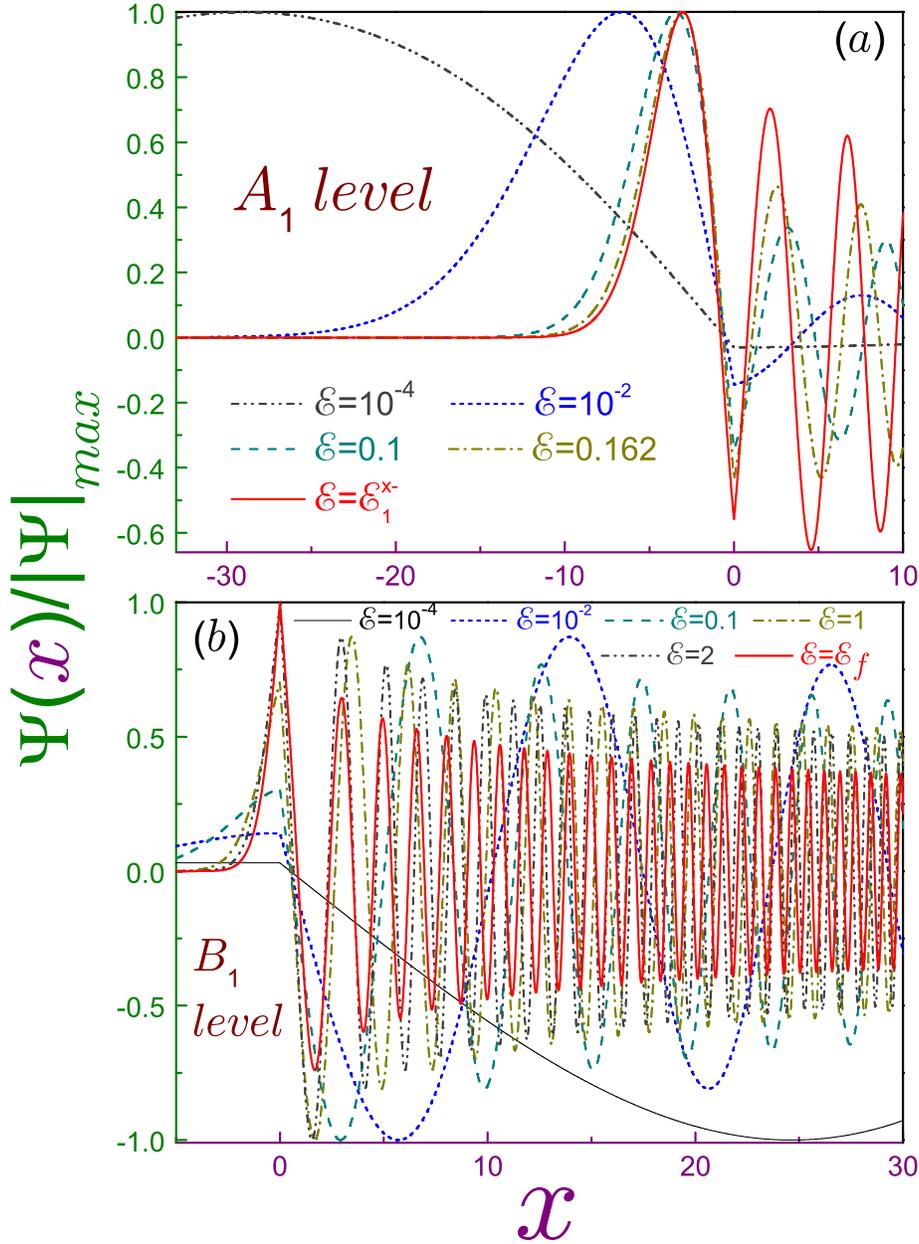}
\caption{\label{DeltaFunction1and2}
The waveforms $\Psi(x)$ of the lowest $A$ [panel (a)] and $B$ [panel (b)] levels at several electric fields $\mathscr{E}$, normalized to the maximum of their absolute values. In the upper plot, the dash-dot-dotted line represents $\mathscr{E}=10^{-4}$, the dotted curve represents $\mathscr{E}=10^{-2}$, the dashed curve represents $\mathscr{E}=0.1$, the dash-dotted curve represents $\mathscr{E}=0.162$, which corresponds to the maximum value of the associated polarization (see Fig.~\ref{DeltaPolarizationFig1}), and the solid line describes the wave function $\Psi_{A_1}(x)$ at the coalescence field $\mathscr{E}_1^{\times-}=0.20773$. For panel (b) the thin solid line represents $\mathscr{E}=10^{-4}$, dotted and dashed curves denote the same voltages as in the upper part, the dash-dotted line denotes $\mathscr{E}=1$, the dash-dot-dotted line denotes $\mathscr{E}=2$, and the thick solid line corresponds to the fundamental field $\mathscr{E}_f$. Note different $x$ and $\Psi$ ranges for each of the panels.
}
\end{figure}

A clear manifestation of the electric breakdown of the QW is revealed by the analysis of the polarization $P$ that, in the coordinate representation, can be written as
\begin{eqnarray}
P_n^\delta(\mathscr{E})=\frac{{\rm Bi}^2\!\left(\!-\frac{E_n}{\mathscr{E}^{2/3}}\right)\!\!\int_{-\infty}^0\!x{\rm Ai}^2\!\left(-\mathscr{E}^{1/3}x\!-\!\!\frac{E_n}{\mathscr{E}^{2/3}}\right)\!dx}{{\rm Bi}^2\!\left(\!-\frac{E_n}{\mathscr{E}^{2/3}}\right)\!\!\int_{-\infty}^0\!{\rm Ai}^2\!\left(-\mathscr{E}^{1/3}x\!-\!\!\frac{E_n}{\mathscr{E}^{2/3}}\right)\!dx}\nonumber\\
\label{DeltaPolarization1}
\frac{+\!{\rm Ai}^2\!\left(\!-\frac{E_n}{\mathscr{E}^{2/3}}\right)\!\!\int_0^\infty\!x{\rm Bi}^2\!\left(-\mathscr{E}^{1/3}x\!-\!\!\frac{E_n}{\mathscr{E}^{2/3}}\right)\!dx}{+\!{\rm Ai}^2\!\left(\!-\frac{E_n}{\mathscr{E}^{2/3}}\right)\!\!\int_0^\infty\!{\rm Bi}^2\!\left(-\mathscr{E}^{1/3}x\!-\!\!\frac{E_n}{\mathscr{E}^{2/3}}\right)\!dx}.
\end{eqnarray}
Note that the primitives in Eq.~\eqref{DeltaPolarization1} can be readily calculated analytically \cite{Vallee1} but applying the limits of integration to the second terms in the numerator and denominator leads to their divergence. Scattering theory has developed special regularization procedures for treating such integrals \cite{Kukulin1} that go back to early efforts in the 1960s \cite{Zeldovich1,Berggren1}. However, to avoid handling of the divergent integrals and their subsequent division in our particular case, it is much easier to use another method for finding the dipole moment that employs the Hellmann-Feynman theorem, which on application to the Hamiltonian from Eq.~\eqref{Hamiltonian1} is
\begin{equation}\label{HellmannFeynman1}
\frac{dE_n}{d\mathscr{E}}=\left\langle\frac{\partial\hat{H}}{\partial\mathscr{E}}\right\rangle=-\left\langle x\right\rangle.
\end{equation}
This immediately leads to the following reworking of Eq.~\eqref{Polarization1} \cite{Olendski1,Moyer3}:
\begin{equation}\label{Polarization3}
P_n(\mathscr{E})=-\frac{dE_n}{d\mathscr{E}}-\left\langle x\right\rangle_{\mathscr{E}=0}.
\end{equation}
The field-free $\delta$-potential is symmetric with respect to the change $x\rightarrow-x$, which results in $\left\langle x\right\rangle_{\mathscr{E}=0}^\delta\equiv0$. Applying the rule of differentiation of the implicit functions to Eq.~\eqref{DeltaPotentialEigenEquation2} leads to
\begin{equation}\label{DeltaPolarization2}
P_n^\delta(\mathscr{E})=-\frac{1}{6}\left[4\frac{E_n}{\mathscr{E}}-\frac{1}{\pi}\frac{1}{f'\left(-E_n/\mathscr{E}^{2/3}\right)}\right].
\end{equation}
The implementation of Eq.~\eqref{Polarization3} to the ground state weak-field limit from Eq.~\eqref{DeltaEnergiesSmallFields1B0} shows that in this regime, the dipole moment increases linearly with applied voltage
\begin{equation}\label{DeltaPolarization3}
P_0^{\delta-}(\mathscr{E})=\frac{5}{8}\,\mathscr{E},\quad\mathscr{E}\ll1,
\end{equation}
meaning that the electron moves in the direction of the external force acting upon it. However, close to the fundamental critical field $\mathscr{E}_f$, the negative divergence of the ground state polarization is gained from Eqs.~\eqref{Polarization3} and \eqref{AsymptotCriticalField1}
\begin{equation}\label{DeltaPolarization4}
P_0^{\delta-}(\mathscr{E})=-\frac{3^{1/2}}{6}\frac{\mathscr{E}_f^{1/2}}{\left(\mathscr{E}_f-\mathscr{E}\right)^{1/2}},\quad\mathscr{E}\rightarrow\mathscr{E}_f.
\end{equation}
To understand this reversal of polarization, the structure of the associated wave functions needs to be considered. Starting from the $A$ levels, the behavior in the very weak fields to the left of the potential is determined by the Airy functions entering the first integrands in Eq.~\eqref{DeltaPolarization1}, which together with the asymptotes from Eq.~\eqref{DeltaEnergiesSmallFields1An} yields:
\begin{equation}\label{DeltaFunctionA1}
\Psi_{A_n}^\delta(x)={\rm Bi}(a_n){\rm Ai}\!\left(-\mathscr{E}^{1/3}x+a_n-\frac{\mathscr{E}^{1/3}}{2}\right),\,x\leq0,\,\mathscr{E}\ll1.
\end{equation}
Applying the properties of the Airy functions \cite{Abramowitz1,Vallee1}, it is found that this waveform has $n$ extrema that are located at
\begin{equation}\label{DeltaExtremaA1}
x_{A_{nm}}^{ext}=\frac{a_n-a_m'}{\mathscr{E}^{1/3}}-\frac{1}{2},\quad\mathscr{E}\ll1,\quad m=1,2,\ldots,n
\end{equation}
[the negative $a_n'$ is the $n$th zero of the derivative of the Ai Airy function, ${\rm Ai}'(a_n')=0$], and the value of the functions at these points is:
\begin{equation}\label{DeltaFunctionA2}
\Psi_{A_n}^\delta\!\left(x_{A_{nm}}^{ext}\right)={\rm Bi}(a_n){\rm Ai}(a_m').
\end{equation}
The wave function to the right possesses an infinite number of fading oscillations, whose amplitudes at low voltages are $\mathscr{E}^{-1/3}$ times smaller than their counterpart(s) at $x<0$:
\begin{eqnarray}
\Psi_{A_n}^\delta(x)=-\frac{1}{2}\,\mathscr{E}^{1/3}{\rm Ai}'(a_n){\rm Bi}\!\left(-\mathscr{E}^{1/3}x+a_n-\frac{\mathscr{E}^{1/3}}{2}\right),\nonumber\\
\label{DeltaFunctionA3}x\geq0,\quad\mathscr{E}\ll1.
\end{eqnarray}
Hence, at extremely weak fields, the $A$ function is located far to the left, which is in accordance with the corresponding negatively diverging polarization derived from Eqs.~\eqref{Polarization3} and \eqref{DeltaEnergiesSmallFields1An}:
\begin{equation}\label{DeltaPolarizationAlimit1}
P_{A_n}^\delta(\mathscr{E})=\frac{2}{3}\frac{a_n}{\mathscr{E}^{1/3}},\quad\mathscr{E}\ll1.
\end{equation}
Note that with the increase of the small voltage the particle, as follows from Eq.~\eqref{DeltaExtremaA1}, moves to the right: this results in the growth of the dipole moment from Eq.~\eqref{DeltaPolarizationAlimit1} and agrees with the electron behavior in the electric field. This shift in the positive $x$ direction is clearly seen in panel (a) of Fig.~\ref{DeltaFunction1and2}. On the contrary, the $B$ level under the same assumption of the small electric intensities resides mainly to the right of the $\delta$-potential with its wave function
\begin{eqnarray}
\Psi_{B_n}^\delta(x)={\rm Ai}(b_n){\rm Bi}\!\left(-\mathscr{E}^{1/3}x+b_n+\frac{\mathscr{E}^{1/3}}{2}\right),\nonumber\\
\label{DeltaFunctionB1}x\geq0,\quad\mathscr{E}\ll1
\end{eqnarray}
exhibiting an infinite number of peaks and dips with the following values
\begin{equation}\label{DeltaFunctionB2}
\Psi_{B_n}^\delta\!\left(x_{B_{nm}}^{ext}\right)={\rm Ai}\left(b_n\right){\rm Bi}\left(b_m'\right)
\end{equation}
[negative coefficients $b_n'$ form an infinite set of roots of the derivative of the Bi Airy function, ${\rm Bi}'(b_n')=0$] located at
\begin{equation}\label{DeltaExtremaB1}
x_{B_{nm}}^{ext}=\frac{b_n-b_m'}{\mathscr{E}^{1/3}}+\frac{1}{2},\quad\mathscr{E}\ll1,\quad m=n,n+1,\ldots.
\end{equation}
Observe that the period of swinging and the distance between extrema increase for decreasing electric intensity. Oscillations to the left, if they exist, are characterized by the amplitudes that are $\mathscr{E}^{-1/3}$ times smaller, after which the waveform exponentially decays with $x$ tending to negative infinity:
\begin{eqnarray}
\Psi_{B_n}^\delta(x)=\frac{1}{2}\,\mathscr{E}^{1/3}{\rm Bi}'(b_n){\rm Ai}\!\left(-\mathscr{E}^{1/3}x+b_n+\frac{\mathscr{E}^{1/3}}{2}\right),\nonumber\\
\label{DeltaFunctionB3}x\leq0,\quad\mathscr{E}\ll1.
\end{eqnarray}
The most important properties to deduce from Eq.~\eqref{DeltaExtremaB1} are: i) at extremely weak fields the particle is located far to the right and ii) with the increase of the (still small) voltage it moves to the {\em left}, which is the opposite direction compared to the $A$ state, and which has just been associated with the electron. This can only be possible if the charge of particle dwelling at the $B$ level is opposite that of its $A$ counterpart. In this way, the natural conclusion is that the $B$ levels correspond to the hole states carrying positive charge. This explains the growth of their energies at small intensities, Eq.~\eqref{DeltaEnergiesSmallFields1Bn}; namely, as the particle moves into the area of the higher potential, its energy increases correspondingly. The general definition of the polarization, Eq.~\eqref{Polarization3}, has to be amended to take into account the different charges of the electrons and holes:
\begin{subequations}\label{PolarizationAmended1}
\begin{eqnarray}\label{PolarizationAmended1A1}
P_{A_n}^\delta(\mathscr{E})&=&-\frac{dE_{A_n}^\delta}{d\mathscr{E}}\\
\label{PolarizationAmended1B1}
P_{B_n}^\delta(\mathscr{E})&=&\frac{dE_{B_n}^\delta}{d\mathscr{E}}.
\end{eqnarray}
\end{subequations}
The last formula together with Eq.~\eqref{DeltaEnergiesSmallFields1Bn} results in positively diverging dipole moments at the weak fields:
\begin{equation}\label{DeltaPolarizationBlimit1}
P_{B_n}^\delta(\mathscr{E})=-\frac{2}{3}\frac{b_n}{\mathscr{E}^{1/3}},\quad\mathscr{E}\ll1,
\end{equation}
conforming to our earlier conclusion regarding the hole locations in this electric regime.  Fig.~\ref{DeltaFunction1and2}(b) exemplifies the wave function shift in the left-hand direction for the lowest $B$ state at small $\mathscr{E}$. Note that in this regime, the $A$ and $B$ waveforms do not exhibit a mutual mirror symmetry with respect to $x=0$ since, as discussed above, the former (latter) is characterized at $x<0$ ($x>0$) by a finite (infinite) number of extrema and with the distance from the origin growing fades as $e^{-\mathscr{E}^{1/2}|x|^{3/2}}$ $\left(\mathscr{E}^{-1/12}x^{-1/4}\right)$.

\begin{figure}
\centering
\includegraphics[width=\columnwidth]{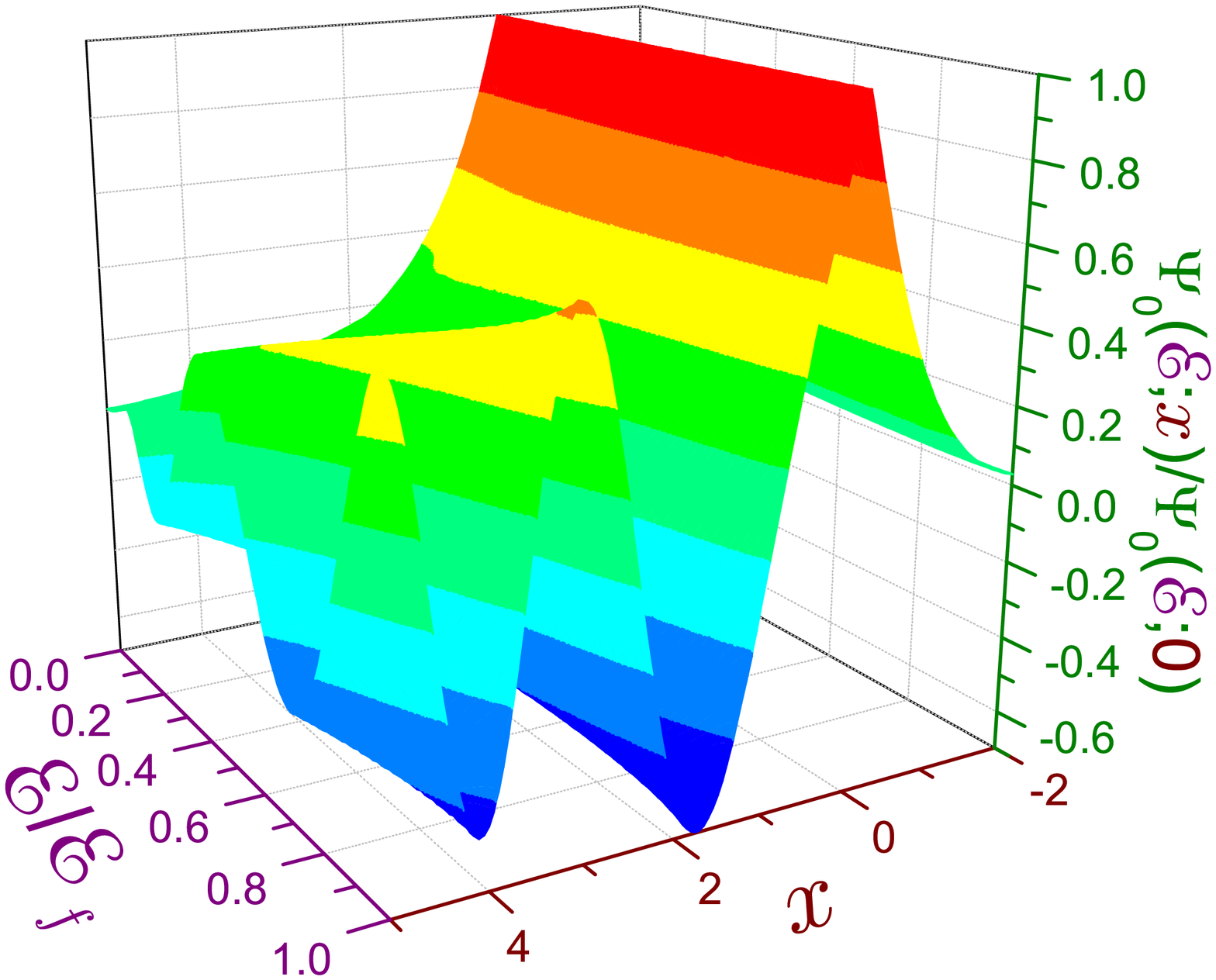}
\caption{\label{DeltaFunction0_3D}
Evolution of the wave function $\Psi_0(x)$ of the lowest quasibound state (normalized to its value at $x=0$) with the electric field $\mathscr{E}$ (in units of $\mathscr{E}_f$).}
\end{figure}

It has thus been shown that, while the particle localization at $E>0$ is completely undetermined for the flat geometry with a short-range potential, the vanishingly weak electric intensity creates electron- and hole-like REQB states in the positive energy continuum that are split in opposite directions. It is important to stress that we use the word "hole" just to underline that the corresponding level behaves like the particle with the positive charge. However, this interesting analogy has its limitations since in the considered system there is no positive charge whatever while in semiconductors the hole carries it because its host atom has a missing electron. From this point of view, it is better to use for our configuration the terms "hole-like state" or quasi-hole, what is tacitly assumed below.

\begin{figure}
\centering
\includegraphics[width=\columnwidth]{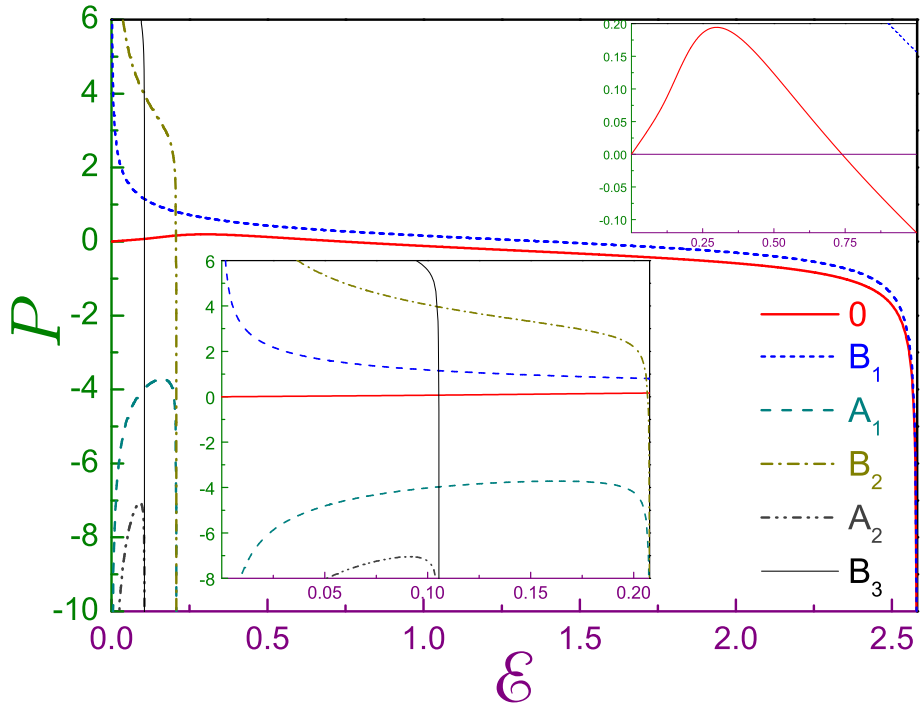}
\caption{\label{DeltaPolarizationFig1}
Polarizations $P$ of the quasibound states of the attractive $\delta$-potential, Eq.~\eqref{DeltaPolarization2}, as a function of the electric field $\mathscr{E}$ where the thick solid line describes the evolution of the zero-field bound state, dotted, dash-dotted, and thin solid lines represent the $B$ levels with $n=1$, $n=2$, and $n=3$, respectively, while the dashed and dash-dot-dotted curves represent the $A$ states with the quantum numbers $n=1$ and $2$, respectively. The upper inset enlarges the view of the ground state polarization at the weak electric intensities and the lower panel depicts the dipole moments in the fields up to $\mathscr{E}_1^{\times-}$.}
\end{figure}

Spatial electron-hole separation $\Delta x_n^{e-h}$ can be defined as the distance between their nearest largest extrema, which for the low voltages reduces to
\begin{equation}\label{DeltaXseparation1}
\Delta x_n^{e-h}=\frac{b_n+a_n'-b_n'-a_n}{\mathscr{E}^{1/3}}+1,\quad\mathscr{E}\ll1.
\end{equation}
This length decreases for larger quantum numbers:
\begin{equation}\label{DeltaXseparation2}
\Delta x_n^{e-h}=\left(\frac{2\pi^2}{3n\mathscr{E}}\right)^{1/3}+1,\quad\mathscr{E}\ll1,\quad n\gg1.
\end{equation}
These equations, together with Fig.~\ref{DeltaFunction1and2}, manifest that at weak electric intensities the electron and hole states are well separated and, accordingly, barely affect each other. The growing field decreases the partition and, as a result, their mutual distortion increases with the corresponding changing shape of the wave functions. The same remains true for the interaction of the ground state with its closest $B$ counterpart. Evolution with the field of the ground level wave function is shown in Fig.~\ref{DeltaFunction0_3D}. At very low voltages, it exhibits fading trigonometric oscillations with the largest amplitude of $\sim\mathscr{E}^{1/6}$ after $x\gtrsim1/\mathscr{E}$ only; accordingly, the associated polarization $P_0$, which, together with its counterparts for several higher lying quasibound states, is shown in Fig.~\ref{DeltaPolarizationFig1}, is determined by the redistribution of the charges near $x=0$, which results in the linear dependence from Eq.~\eqref{DeltaPolarization3}. At the same time, the nearest $B_1$ state is rapidly moving to the left, causing the steep decrease in the dipole moment $P_{B_1}$ seen in Fig.~\ref{DeltaPolarizationFig1}. By pushing the two states closer to each other, the increasing electric intensity forces them to interact more strongly with the concomitant larger deformations of the wave functions that exhibit higher frequencies of oscillations. This increase in the voltage leading to mutual interference of the electron- and hole-like states, which can be construed as an attraction of opposite charges, also affects their energies, compelling them to change direction and move towards each other, which, according to Eq.~\eqref{Polarization3}, influences the ground dipole moment in such a way that at $\mathscr{E}=0.299$ it passes through the maximum of $P_{0_{max}}^{\delta-}=0.1943$, after which it decreases and moves closer and closer to its $B_1$ partner. As the field approaches the coalescence value, the squeezing of the waveforms gets stronger, which leads to a drastic decrease in the polarizations. Simultaneously, there is an increasing degree of similarity between the two levels and, at the merger, one has two identical solutions with the same energies and wave functions. This collision of levels at the coalescence field can be considered as an electron-hole recombination and, as mentioned above, the voltage $\mathscr{E}_f$ is the breakdown field beyond which the $\delta$-potential cannot bind the electron \cite{Moyer3,Moyer4}. However, the existence at $\mathscr{E}>\mathscr{E}_f$ of the complex conjugate solutions of Eq.~\eqref{DeltaPotentialEigenEquation2} suggests another interpretation; namely, a creation at these electric intensities of a composite particle when the electron and hole stick together, forming an exciton; their individual motions cannot be considered independently since they correlate with each other. Note that we have considered the one-particle Schr\"{o}dinger equation [see Eqs.~\eqref{Schrodinger1} and \eqref{Hamiltonian1}] but even in this simplest form it hints at the emergence of the many-particle phenomena in the electric field. More advanced theories should be employed to address this issue, which lies beyond the scope of the present research. Of course, what has been said above applies also to the higher lying states for which the coalescence fields are lower due to the smaller electron-hole separation, see Eq.~\eqref{DeltaXseparation2}. At the end of this paragraph, we will point out that a comparison of the corresponding wave functions from Figs.~\ref{DeltaPotentialComplexEnergyFunction0}(b) and \ref{DeltaFunction0_3D} vividly underlines the difference between the two approaches.

Under some conditions specified below, {\em real} energies $E_n$, which are solutions of Eq.~\eqref{DeltaPotentialEigenEquation2}, can be supplemented by the associated half widths $\Gamma_n^{BW}$, and the set of these two quantities can be construed again as the {\em complex} energy; namely, applying a Taylor expansion to the real part of the numerator and denominator on the right-hand side of Eq.~\eqref{DeltaPotentialScatteringMatrix1} around its zeros, the expression for the scattering matrix can be recast in the well-known Breit-Wigner form \cite{Newton1,Landau1,Baz1,Kukulin1,Breit1}
\begin{subequations}\label{Auxiliary1}
\begin{align}\label{BreitWigner1}
S(E)&=e^{i2\varphi_P}\frac{E-E_n-i\left.\Gamma_n^{BW}\!\right/2}{E-E_n+i\left.\Gamma_n^{BW}\!\right/2},
\intertext{
where a slowly varying potential phase $\varphi_P$ takes, for the current geometry, a constant value of $\pi/2$ resulting in the following expression for the scattering probability:}
\label{ScatProbab2}
p(E)&=\frac{\left(\left.\Gamma_n^{BW}\!\!\right/2\right)^2}{\left(E-E_n\right)^2+\left(\left.\Gamma_n^{BW}\!\!\right/2\right)^2}.
\intertext{These equations are valid at}
\label{Condition1}
\left|\Gamma_n^{BW}\right|&\ll1,\,\left|E-E_n\right|\!\ll\!\left|E_n-E_{n+1}\right|,\,\left|E-E_n\right|\!\ll\!\left|E_n-E_{n-1}\right|.
\end{align}
\end{subequations}
The Breit-Wigner half width $\Gamma_n^{BW}$, which, according to Eq.~\eqref{ScatProbab2}, defines a full width at half maximum (FWHM) of the scattering probability $p(\mathscr{E};E)$, is given by
\begin{equation}\label{DeltaPotentialHalfWidth1}
\Gamma_n^{BW}=-2\mathscr{E}^{2/3}\frac{{\rm Ai}^2\!\left(\!\left.-E_n\!\right/\!\mathscr{E}^{2/3}\right)}{f'\!\left(\!\left.-E_n\!\right/\!\mathscr{E}^{2/3}\right)}.
\end{equation}

For the weak fields, $\mathscr{E}\ll1$,
\begin{subequations}\label{DeltaHalfWidthSmallFields1}
\begin{eqnarray}\label{DeltaHalfWidthSmallFields1B0}
\Gamma_0^{BW}&=&2\exp\!\left(\!-\frac{4}{3}\frac{1}{\mathscr{E}}\right)\\
\label{DeltaHalfWidthSmallFields1An}
\Gamma_{A_n^\pm}^{BW}&=&\frac{1}{2\pi{\rm Bi}^2(a_n)}\,\mathscr{E}^{4/3}\\
\label{DeltaHalfWidthSmallFields1Bn}
\Gamma_{B_n^\pm}^{BW}&=&-2\frac{{\rm Ai}(b_n)}{{\rm Bi}'(b_n)}\,\mathscr{E}^{2/3}.
\end{eqnarray}
\end{subequations}
First, again note, similar to Eqs.~\eqref{DeltaComplexSolutionsTwoSets1}, the different powers of the field for the $A$ and $B$ resonances. This is explained by the fact that the corresponding waveforms are not exact replicas of each other under the transformation $x\rightarrow-x$. Next, Eqs.~\eqref{DeltaEnergiesSmallFields1B0} and \eqref{DeltaHalfWidthSmallFields1B0} are, obviously, just the expression for the complex energy from Eq.~\eqref{DeltaPotentialAsymptotics1} derived by the Gamow-Siegert method, while Eqs.~\eqref{DeltaEnergiesSmallFields1An} and \eqref{DeltaHalfWidthSmallFields1An} compose complex solutions from Eq.~\eqref{DeltaComplexSolutionsTwoSets1_Set1} that, accordingly, describe the states located at the far left (not taking into account, of course, the divergence at large positive $x$). However, real and imaginary parts of the second set, Eq.~\eqref{DeltaComplexSolutionsTwoSets1_Set2}, of the complex solutions of Eq.~\eqref{DeltaPotentialEigenEquation1} bear no resemblance to the corresponding numbers from Eqs.~\eqref{DeltaEnergiesSmallFields1Bn} and \eqref{DeltaHalfWidthSmallFields1Bn}. Moreover, the results that coincided at small electric intensities diverge considerably from each other at stronger $\mathscr{E}$. Mathematically, this is due to the fact that {\em complex} solutions of Eq.~\eqref{DeltaPotentialEigenEquation1} are, in general, different from the complex numbers whose {\em real} parts are determined from Eq.~\eqref{DeltaPotentialEigenEquation2} and whose {\em imaginary} components obey Eq.~\eqref{DeltaPotentialHalfWidth1}.

Eqs.~\eqref{DeltaHalfWidthSmallFields1An} and \eqref{DeltaHalfWidthSmallFields1Bn} can be further simplified for large quantum numbers \cite{Alvarez1} when the approximate analytic expressions for the roots $a_n$ and $b_n$ do exist \cite{Abramowitz1,Vallee1}:
\begin{subequations}\label{AiryZeros1}
\begin{eqnarray}\label{AiryAiZeros1}
a_n=-\left[\frac{3\pi}{8}(4n-1)\right]^{2/3},\quad n\gg1\\
\label{AiryBiZeros1}
b_n=-\left[\frac{3\pi}{8}(4n-3)\right]^{2/3},\quad n\gg1.
\end{eqnarray}
\end{subequations}
Even though these asymptotic formulae were derived for large $n$, they also provide reasonably good accuracy for small $n$. For example, in the extreme opposite limit of $n=1$ the exact values are \cite{Abramowitz1} $a_1=-2.3381$ and $b_1=-1.1737$ while their approximations from Eqs.~\eqref{AiryZeros1} yield $-2.3203$ and $-1.1155$, respectively. These expressions also lead to a drastic reduction of the products and ratio of the Airy functions in the above equations:
\begin{subequations}\label{DeltaEnergiesSmallFieldsLargeN1}
\begin{eqnarray}
\label{DeltaEnergiesSmallFieldsLargeN1An}
E_n^{A\pm}&=&\left[\frac{3\pi}{8}(4n-1)\mathscr{E}\right]^{2/3}\mp\frac{1}{2}\,\mathscr{E},\quad\mathscr{E}\ll1,\,n\gg1\\
\label{DeltaEnergiesSmallFieldsLargeN1Bn}
E_n^{B\pm}&=&\left[\frac{3\pi}{8}(4n-3)\mathscr{E}\right]^{2/3}\pm\frac{1}{2}\,\mathscr{E},\quad\mathscr{E}\ll1,\,n\gg1\\
\label{DeltaHalfWidthSmallFieldsLargeN1An}
\Gamma_{A_n^\pm}^{BW}&=&\frac{1}{2}\left[\frac{3\pi}{8}(4n-1)\right]^{1/3}\mathscr{E}^{4/3},\quad\mathscr{E}\ll1,\,n\gg1\\
\label{DeltaHalfWidthSmallFieldsLargeN1Bn}
\Gamma_{B_n^\pm}^{BW}&=&-2\left[\frac{3\pi}{8}(4n-3)\right]^{-1/3}\mathscr{E}^{2/3},\quad\mathscr{E}\ll1,\,n\gg1.
\end{eqnarray}
\end{subequations}
The first thing to notice is that the half widths of the $B$ resonances are negative; this superficially contradicts a commonly accepted view regarding the positiveness of $\Gamma$. However, in the wake of the previous discussion in this subsection, this result is not surprising as it means that the corresponding states leak in opposite directions: at low voltage the $A$ level is localized mainly to the left of the $\delta$-potential and as the field grows it tries to expand to its right, which is considered to be a positive route with the same sign of the half width. On the other hand, the $B$ orbital, which at $\mathscr{E}\ll1$ is found mainly at $x\gg1$, tries to make its way to the left of the potential. This opposing direction of the leakage is mathematically reflected in the negativeness of the factor $\Gamma$. Note that the scattering probability $p$ from Eq.~\eqref{ScatProbab2} depends on the square of the half width what means that it can be experimentally measured for $\Gamma_n^{BW}<0$. It is instructive to draw parallels with the Gamow-Siegert states, where the opposite signs of the half widths describe the  behavior of the {\em same} level at earlier and later times, as discussed in Sec.~\ref{GamowSiegertStates1}; however, in the time-independent Breit-Wigner picture they are associated with different localizations of the {\em two} states. Observe also that the Gamow half widths are the imaginary parts of the complex conjugate solutions of Eq.~\eqref{DeltaPotentialEigenEquation1}; therefore, their absolute values are equal to each other, whereas for the Breit-Wigner configuration the magnitudes are different, as Eqs.~\eqref{DeltaHalfWidthSmallFieldsLargeN1An} and \eqref{DeltaHalfWidthSmallFieldsLargeN1Bn} exemplify.

With the growth of the field, the amplitudes of the half widths increase too, narrowing in this way the conditions of applicability of the Breit-Wigner approximation from Eq.~\eqref{Condition1}; in particular, and in a similar way to the derivation of Eq.~\eqref{AsymptotCriticalField1}, it can be shown that in the near vicinity of the fundamental critical field $\mathscr{E}_f$ the half width dependencies on the applied voltage are:
\begin{equation}\label{AsymptotCriticalField2}
\Gamma_{0,B_1}^{BW}=\pm\frac{\mathscr{E}_f^{3/2}}{\left(\mathscr{E}_f-\mathscr{E}\right)^{1/2}},\quad\mathscr{E}\rightarrow\mathscr{E}_f.
\end{equation}
For the smaller critical fields the divergence formula can be derived in a similar way to Eq.~\eqref{AsymptotCriticalField1_1}. Of course, for such wide resonances the Breit-Wigner approximation can no longer be used and instead the original exact equation~\eqref{DeltaPotentialScatteringMatrix1} should be applied.

As a final part of this subsection, let us point out that coalescences characterized by the fields $\mathscr{E}_n^{\times\pm}$ and energies $E_n^{\times\pm}$ present a special case of so-called exceptional points (EPs) \cite{Heiss2} where a merging of the two (or more) levels with the variation of some physical parameter is governed by the square root singularities from Eqs.~\eqref{AsymptotCriticalField1} or \eqref{AsymptotCriticalField1_1}. Contrary to usual degeneracy, EP exhibits not only equal energies of the different states but also linear dependent eigenfunctions, as discussed above. These spectral singularities, which are ubiquitous in nature, produce dramatic effects in, e.g., multichannel scattering, anomalous time behavior \cite{Heiss3}, etc.; for instance, for our geometry, EPs are characterized by the diverging polarizations, as exemplified by Eq.~\eqref{DeltaPolarization4} and Fig.~\ref{DeltaPolarizationFig1}. Note that, in general, the physical parameter variation of which leads to the coalescence is {\em complex} with the corresponding {\em complex} eigenvalues \cite{Heiss2} while for the $\delta$-potential it is the electric field with the {\em real} magnitude that causes a merging of the two REQB levels with {\em real} energies and their subsequent motion into the {\em complex} plane when the voltage is increased.

\subsection{Time Delay}
The expression for the time delay of the $\delta$-potential in the electric field is:
\begin{eqnarray}
\tau_W^{\delta\pm}&=&-\frac{8\pi}{\mathscr{E}^{2/3}}\nonumber\\
\label{DeltaTimeDelay1}
&\times&\frac{{\rm Ai}_0\left({\rm Ai}_0\mp\mathscr{E}^{1/3}{\rm Ai}_0'\right)}{4\pi^2{\rm Ai}_0^2\left({\rm Ai}_0^2+{\rm Bi}_0^2\right)\pm4\pi\mathscr{E}^{1/3}{\rm Ai}_0{\rm Bi}_0+\mathscr{E}^{2/3}}.
\end{eqnarray}
This formula, where, for brevity, the subscript '$0$' at each of the functions means that they are evaluated at the value of $\eta_0$ defined above [${\rm Ai}_0\equiv{\rm Ai}\left(-E/\mathscr{E}^{2/3}\right)$, etc.], was derived from the corresponding counterpart for the phase $\varphi_S$
\begin{equation}\label{DeltaPhase1}
\varphi_S^{\delta\pm}=\pi+\arctan\!\!\left(4\frac{{\rm Ai}_0^2\left(2{\rm Ai}_0{\rm Bi}_0\pm\mathscr{E}^{1/3}/\pi\right)}{4{\rm Ai}_0^4-\left(2{\rm Ai}_0{\rm Bi}_0\pm\mathscr{E}^{1/3}/\pi\right)^2}\right)
\end{equation}
with the use of the properties of the Airy functions. For large negative energies, the delay time very abruptly approaches zero:
\begin{subequations}\label{DeltaTimeDelayAsymptotics1}
\begin{align}\label{DeltaTimeDelayAsymptotics1_NegativeEnergy}
\tau_W^{\delta\pm}=\mp\frac{2}{\mathscr{E}}\exp\!\left(-\frac{4}{3}\frac{|E|^{3/2}}{\mathscr{E}}\right),\quad E\ll-1,
\intertext{which physically means that the incident particle at such energies does not 'see' the $\delta$-potential and is reflected from the tilted potential almost immediately \cite{Emmanouilidou2}. In the opposite limit the Wigner time degenerates to}
\label{DeltaTimeDelayAsymptotics1_PositiveEnergy}
\tau_W^{\delta\pm}=\mp\frac{4}{\mathscr{E}}\cos\!\left(\frac{4}{3}\frac{E^{3/2}}{\mathscr{E}}\right),\quad E\gg1,
\end{align}
\end{subequations}
meaning that it reaches its maxima
\begin{equation}\label{AsymptotDeltaStrongField1}
\tau_{max_n}^{\delta\pm}=\frac{4}{\mathscr{E}}
\end{equation}
at
\begin{subequations}\label{DeltaTimeDelayAsymptotics3}
\begin{eqnarray}\label{DeltaTimeDelayAsymptotics3Plus}
E^{\delta+}_{max_n}&=&\left[\frac{3\pi}{4}\left(2n+1\right)\mathscr{E}\right]^{2/3}\\
\label{DeltaTimeDelayAsymptotics3Minus}
E^{\delta-}_{max_n}&=&\left(\frac{3\pi}{2}n\mathscr{E}\right)^{2/3},
\end{eqnarray}
\end{subequations}
where the non-negative integer $n$ and the field $\mathscr{E}$ are such that the condition $E\gg1$ is satisfied basically reducing it to the requirements $\mathscr{E}\gg1$ or/and $n\gg1$. This oscillating behavior is explained by the interference of the incident and reflected waves in the region to the left of the well where, as stated above, the interplay of the electric field and $\delta$-potential creates a triangular QW with a transparency of its right wall depending on the field and energy. This interaction of the incoming and outgoing fluxes also explains the negative delay times that directly follow from Eq.~\eqref{DeltaTimeDelayAsymptotics1_PositiveEnergy}; namely, at some particular energies the interference makes the right wall of the triangular QW an impenetrable surface, meaning that the electron is reflected from the point $x=0$ and not from $x=-E/\mathscr{E}$, which is a quasi-classical turning point without the $\delta$-potential. Note that the distance between the maxima of the Wigner times increases with voltage while their peak values are inversely proportional to it.

\begin{figure}
\centering
\includegraphics[width=\columnwidth]{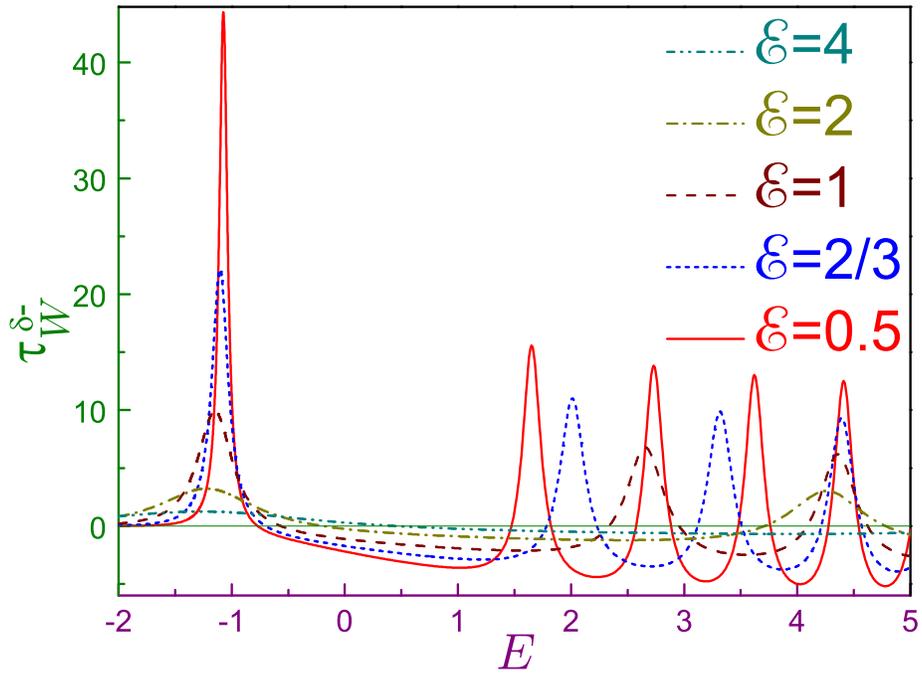}
\caption{\label{DeltaWellWignerTime}
Wigner delay time $\tau_W^{\delta-}$ of the attractive $\delta$-potential as a function of energy $E$ at several electric fields where the solid line is for $\mathscr{E}=0.5$, dotted -- for $\mathscr{E}=2/3$, dashed -- for $\mathscr{E}=1$, dash-dotted -- for $\mathscr{E}=2$, and dash-dot-dotted curve -- for $\mathscr{E}=4$. Thin horizontal line denotes zero time.}
\end{figure}

Fig.~\ref{DeltaWellWignerTime} shows the delay time as a function of energy at several different electric intensities. At each fixed field an infinite number of maxima can be observed with their sharpness and peak value being field-dependent. Accordingly, each resonance is characterized by three parameters: the location of the delay time peak, its maximum at this energy, and the corresponding FWHM. For weak fields, it can be shown that the energies of the resonances in the positive part of the spectrum are:
\begin{eqnarray}
E^{\delta\pm}_{max_n}=-a_n\mathscr{E}^{2/3}\mp\frac{1}{2}\,\mathscr{E}-\frac{1}{16}\frac{{\rm Ai}'(a_n)^2{\rm Bi}'(a_n)}{{\rm Bi}^3(a_n)}\,\mathscr{E}^2,\nonumber\\
\label{DeltaResonancesPositiveEnergiesSmallField1}\mathscr{E}\ll1,\quad n\geq1,
\end{eqnarray}
i.e., they are close to the energies of the $A$-type quasibound states from Eq.~\eqref{DeltaEnergiesSmallFields1An}. Physically, this proximity to only the $A$ (and not to the $B$) levels is again explained by the formation to the left of the $\delta$-potential of the triangular QW that captures the electron for some time, while from the mathematical point of view both terms in the denominator of $\arctan$ in Eq.~\eqref{DeltaPhase1} are small and the tiny variation in the energy range close to that from Eq.~\eqref{DeltaResonancesPositiveEnergiesSmallField1} leads to subtle interplay between them, resulting in pronounced resonance with the maximum
\begin{equation}\label{DeltaResonancesPositiveEnergiesSmallField2}
\tau^{\delta\pm}_{max_n}=-8\frac{{\rm Bi}(a_n)}{{\rm Ai}'(a_n)}\frac{1}{\mathscr{E}^{4/3}},\quad\mathscr{E}\ll1,\quad n\geq1,
\end{equation}
which, according to the previous reasoning, for the rather large $n$ can be approximated as:
\begin{equation}\label{DeltaResonancesPositiveEnergiesSmallField3}
\tau^{\delta\pm}_{max_n}=8\left[\frac{3\pi}{8}(4n-1)\right]^{-1/3}\frac{1}{\mathscr{E}^{4/3}},\quad\mathscr{E}\ll1,\,n\gg1.
\end{equation}
Note that the peak decreases with growing $n$ and tends to infinity for the vanishing fields as $\mathscr{E}^{-4/3}$. For $E<0$ and small electric intensities, the resonance location $E^{\delta-}_{max_0}$ is exponentially close to the lowest quasibound state energy from Eq.~\eqref{DeltaEnergiesSmallFields1B0} [or, alternatively, to the negative real part of the Gamow-Siegert energy, Eq.~\eqref{DeltaPotentialAsymptotics1}]; thus, assuming their equality, the phase shift is determined as
\begin{equation}\label{DeltaPhase2}
\varphi_S=\pi-\arctan\frac{\Gamma_0(E-E_0)}{(E-E_0)^2-(\Gamma_0/2)^2},\quad\mathscr{E}\ll1,
\end{equation}
resulting in the Lorentzian shape of the time delay:
\begin{equation}\label{DeltaResonanceNegativeEnergySmallField1}
\tau_0^{\delta-}=\frac{\Gamma_0}{(E-E_0)^2+(\Gamma_0/2)^2},
\end{equation}
whose maximum exponentially approaches infinity with the disappearing fields:
\begin{equation}\label{DeltaResonanceNegativeEnergySmallField2}
\tau_{max_0}^{\delta-}=\frac{4}{\Gamma_0}=2\exp\!\left(\frac{4}{3}\frac{1}{\mathscr{E}}\right),\quad\mathscr{E}\ll1.
\end{equation}
Its FWHM in this regime is equal to $\Gamma_0$ from Eqs.~\eqref{DeltaHalfWidthAsymptotics1} or \eqref{DeltaHalfWidthSmallFields1B0}. In the opposite limit of $\mathscr{E}\gg1$, Eq.~\eqref{DeltaTimeDelayAsymptotics1_PositiveEnergy} gives
\begin{subequations}\label{DeltaResonanceHalfWidth1}
\begin{eqnarray}\label{DeltaResonanceHalfWidth1Minus}
\Gamma_n^{\delta-}=\left(\frac{3\pi}{4}\,\mathscr{E}\right)^{2/3}\left[\left(2n+\frac{1}{2}\right)^{2/3}-\left(2n-\frac{1}{2}\right)^{2/3}\right],\,\mathscr{E}\gg1\\
\label{DeltaResonanceHalfWidth1Plus}
\Gamma_n^{\delta+}=\left(\frac{3\pi}{4}\,\mathscr{E}\right)^{2/3}\left[\left(2n+\frac{3}{2}\right)^{2/3}-\left(2n+\frac{1}{2}\right)^{2/3}\right],\,\mathscr{E}\gg1,
\end{eqnarray}
\end{subequations}
which for large $n$ degenerates to
\begin{equation}\label{DeltaResonanceHalfWidth2}
\Gamma_n^{\delta\pm}=\frac{1}{3}\left(\frac{3\pi}{2}\,\mathscr{E}\right)^{2/3}\frac{1}{n^{1/3}},\quad\mathscr{E}\gg1,\quad n\gg1.
\end{equation}
\begin{figure}
\centering
\includegraphics[width=\columnwidth]{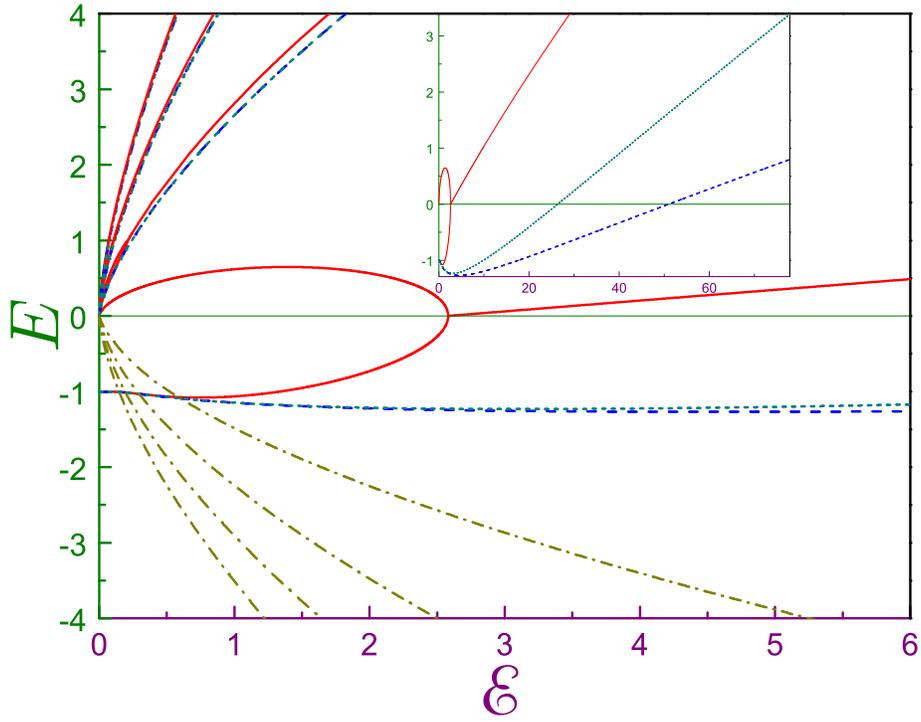}
\caption{\label{DeltaAllEnergies}
Comparison of the different methods for calculating the energies of the attractive $\delta$-potential in the electric field $\mathscr{E}$ where the solid lines show solutions of Eq.~\eqref{DeltaPotentialEigenEquation2}, dashed curves denote locations of the maxima of the Wigner delay time, and the dotted and dash-dotted lines are the real parts of the first and second sets, respectively, of solutions of Eq.~\eqref{DeltaPotentialEigenEquation1}. The inset shows the lowest levels at a different scale.}
\end{figure}

Fig.~\ref{DeltaAllEnergies} compares the energies calculated by the three methods. It can be seen that the positive real parts of the complex Gamow-Siegert energies are indistinguishable from the locations of the time delay maxima with $n\geq1$: due to their closeness, they are not resolved in the figure. However, the lowest maximum position and the corresponding real part of the complex energy, which are the same at weak fields, deviate from each other with increasing voltage; for example, $E^{\delta-}_{max_0}$ reaches a minimum of $-1.2685$ at $\mathscr{E}=4.57$ which, if compared to the analogous data of the other methods detailed above, means that the Wigner time calculations provide the lowest estimate. The divergence between the energies increases for stronger fields, as the inset demonstrates. As emphasized above, this increase with the field of the difference between the outcomes of the two approaches is mathematically explained by the different equations, while the physical reason lies in the fact that Gamow-Siegert solutions describe the outgoing oscillation while the real energy scattering approach operates with the standing waves that do not carry current. For completeness, the plot also shows the evolution of the real components of the second set of solutions of Eq.~\eqref{DeltaPotentialEigenEquation1}, which for the weak fields are described by Eq.~\eqref{DeltaComplexSolutionsTwoSets1_Set2}. Their characteristic feature is the fact that with the increase of the electric intensity they cross the level that evolved from the zero-field bound state \cite{Emmanouilidou2}.

\begin{figure}
\centering
\includegraphics[width=\columnwidth]{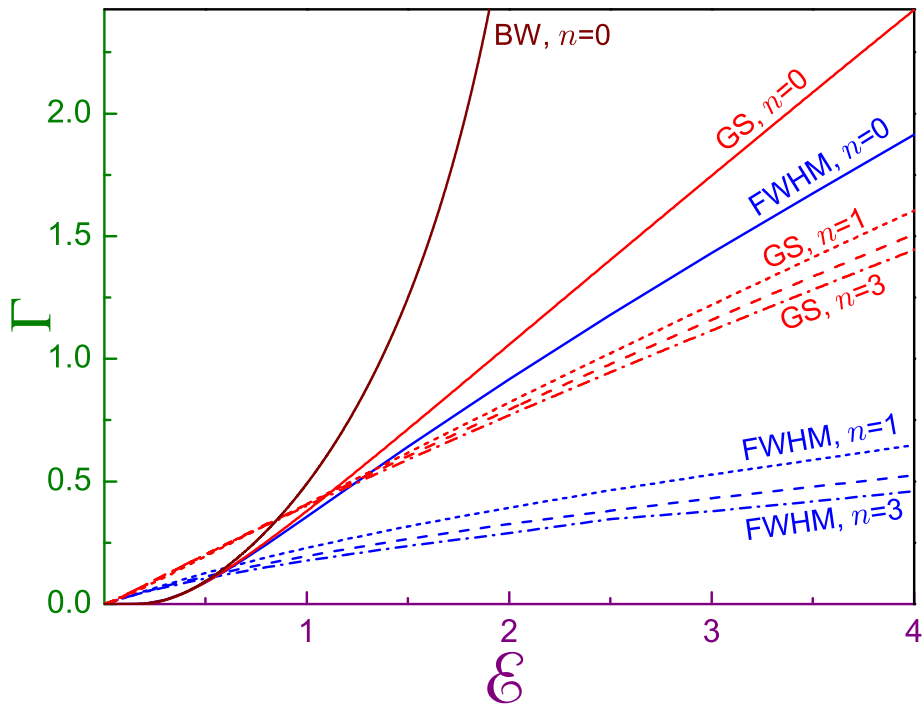}
\caption{\label{DeltaHalfWidthResonances1}
Half widths $\Gamma$ calculated for the Gamow-Siegert (GS) states, and as FWHMs of the Wigner resonance peaks denoted by the corresponding acronyms. For comparison, the $n=0$ curve for the Breit-Wigner (BW) half width, Eq.~\eqref{DeltaPotentialHalfWidth1}, is also shown. Solid lines denote $n=0$, dotted lines are for $n=1$, dashed curves -- $n=2$, and dash-dotted lines denote $n=3$.}
\end{figure}

The variation of the half widths with the field calculated with the help of the complex Airy functions and as FWHMs of the scattering configuration is depicted in Fig.~\ref{DeltaHalfWidthResonances1} where a positive $\Gamma_0^{BW}$ is also shown. Qualitatively, the transformation from the weak field regime described by Eqs.~\eqref{DeltaHalfWidthSmallFields1B0} and \eqref{DeltaHalfWidthSmallFieldsLargeN1An}, when the half widths of the states with the larger $n$ are greater than their counterparts with the smaller quantum numbers, to the high voltage configuration with its dependence from Eq.~\eqref{DeltaResonanceHalfWidth2} is typical for either approach: crossings of the lines for different $n$ are clearly seen in the plot. Quantitatively, the Breit-Wigner approximation produces the larger estimate of $\Gamma$ followed by the Gamow model with increasing difference with the field: the half widths with $n\geq1$ are approximately equal at $\mathscr{E}\lesssim0.02$ only while, for $n=0$, they remain approximately the same up to $\mathscr{E}\sim0.6$. Thus, in general, the complex energy method leads to smaller lifetimes compared to those obtained by the scattering approach with real $E$.

\section{Robin Wall}\label{Sec_Robin}
Consider the motion of the particle on the half line $0\leq x<\infty$ subject at the left edge to the BC \cite{Olendski2,Seba1,Pazma1,Fulop1,Belchev1,Georgiou1}
\begin{equation}\label{Robin2}
\Lambda\Psi'(0)=\Psi(0),
\end{equation}
which is just the 1D analogue of Eq.~\eqref{Robin1}. Well-known particular cases of this general interface demand are the Dirichlet requirement, $\Lambda_D=0$, with the vanishing function at the left end,
\begin{subequations}\label{BCRobin1}
\begin{align}\label{BCRobin1_D}
\Psi_D(0)=0,\\
\intertext{and the Neumann one, $\Lambda_N=\infty$, when its spatial derivative vanishes at the confining surface,}
\label{BCRobin1_N}
\Psi_N'(0)=0.
\end{align}
\end{subequations}
In the absence of the fields, $\mathscr{E}=0$, the BC from Eq.~\eqref{Robin2} is formally identical to Eq.~\eqref{MatchingCondtions1_2}, which allows us to immediately conclude that the Robin wall with a negative extrapolation length acts as an attractive interface, creating a bound state with the energy from Eq.~\eqref{EnergyDeltaZeroFields1} and the wave function 
\begin{equation}\label{FunctionRobinZeroFields1}
\Psi(x)=\left(\frac{2}{|\Lambda|}\right)^{1/2}\exp\!\left(-\frac{x}{|\Lambda|}\right),
\end{equation}
which satisfies the normalization
\begin{equation}\label{Normalization2}
\int_0^\infty\Psi^2(x)dx=1.
\end{equation}
Experimentally, the surfaces with negative de Gennes distances were fabricated with the help of superconductors \cite{Fink1,Kozhevnikov1}. This model also approximates, as the limiting case, the finite continual potentials \cite{Pazma1,Fulop1}, which are created by using thin layers of different types of semiconductors. More relevant references relating to the Robin structures can be found in Refs.~\cite{Olendski3,Olendski4,Olendski5,Grebenkov1}.

The wall is highly asymmetric with respect to the sign of the applied field: for negative electric intensities, $\mathscr{E}<0$, the spectrum is completely discrete, while for positive voltages it stays continuous. The former configuration is discussed elsewhere \cite{Olendski2}. In this section, we will address the case of the electric force that attempts to push the electron away from the wall, $\mathscr{E}>0$. For the finite non-vanishing $\Lambda$ the same dimensionless units as those in Sect.~\ref{Sec_Delta} will be used, while for the Dirichlet or Neumann BC the most appropriate unit of distance is the reduced Compton wavelength $\lambdabar=\hbar/(mc)$, which naturally leads to the units of energy $mc^2$, and electric field -- $m^2c^3/(e\hbar)$, with $c$ being the speed of light. As a result, the equation of motion takes the form 
\begin{equation}\label{Schrodinger3}
-\Psi''(x)-\mathscr{E}x\Psi(x)=E\Psi(x),
\end{equation}
and the BC changes to either
\begin{equation}\label{BCRobin2}
\pm\Psi'(0)=\Psi(0)
\end{equation}
with the sign corresponding to that of the de Gennes distance, or Eqs.~\eqref{BCRobin1}. Below, we will use the superscript D (N) to denote the Dirichlet \cite{Dean1} (Neumann) type of BC at the interface, while the character R followed, if necessary, by the plus (minus) sign will refer to the Robin surface with a positive (negative) extrapolation length. 

The general line of investigation is the same as that used for the $\delta$-potential, but to make the exposition shorter the expression for the scattering matrix $S$ is written from the outset as:
\begin{subequations}\label{RobinScatteringMatrix1}
\begin{eqnarray}
S^{R\pm}(\mathscr{E};E)=\nonumber\\
\label{RobinScatteringMatrix1_R}
\!\!-\!\frac{{\rm Bi}\!\left(\!-\!\frac{E}{\mathscr{E}^{2/3}}\!\!\right)\!\!\pm\!\!\mathscr{E}^{1/3}\!{\rm Bi}'\!\left(\!-\!\frac{E}{\mathscr{E}^{2/3}}\!\!\right)\!\!-\!\!i\!\!\left[{\rm Ai}\!\left(\!-\!\frac{E}{\mathscr{E}^{2/3}}\!\!\right)\!\!\pm\!\!\mathscr{E}^{1/3}\!{\rm Ai}'\!\left(\!-\!\frac{E}{\mathscr{E}^{2/3}}\!\!\right)\!\!\right]}{{\rm Bi}\!\left(\!-\!\frac{E}{\mathscr{E}^{2/3}}\!\!\right)\!\!\pm\!\!\mathscr{E}^{1/3}\!{\rm Bi}'\!\left(\!-\!\frac{E}{\mathscr{E}^{2/3}}\!\!\right)\!\!+\!\!i\!\!\left[{\rm Ai}\!\left(\!-\!\frac{E}{\mathscr{E}^{2/3}}\!\!\right)\!\!\pm\!\!\mathscr{E}^{1/3}\!{\rm Ai}'\!\left(\!-\!\frac{E}{\mathscr{E}^{2/3}}\!\!\right)\!\!\right]}\\
\label{RobinScatteringMatrix1_D}
S^D(\mathscr{E};E)=-\frac{{\rm Bi}\!\left(-\frac{E}{\mathscr{E}^{2/3}}\right)-i{\rm Ai}\!\left(-\frac{E}{\mathscr{E}^{2/3}}\right)}{{\rm Bi}\!\left(-\frac{E}{\mathscr{E}^{2/3}}\right)+i{\rm Ai}\!\left(-\frac{E}{\mathscr{E}^{2/3}}\right)}\\
\label{RobinScatteringMatrix1_N}
S^N(\mathscr{E};E)=-\frac{{\rm Bi}'\!\left(-\frac{E}{\mathscr{E}^{2/3}}\right)-i{\rm Ai}'\!\left(-\frac{E}{\mathscr{E}^{2/3}}\right)}{{\rm Bi}'\!\left(-\frac{E}{\mathscr{E}^{2/3}}\right)+i{\rm Ai}'\!\left(-\frac{E}{\mathscr{E}^{2/3}}\right)}.
\end{eqnarray}
\end{subequations}
A condition of zeroing its denominator produces the equations for determining the complex resonance energies $E_{res}$, which are written in the form:
\begin{subequations}\label{RobinEigenEquation1}
\begin{eqnarray}\label{RobinEigenEquation1_R}
{\rm Ai}\!\left(-\frac{E^{R\pm}}{\mathscr{E}^{2/3}}\,e^{i2\pi/3}\!\right)&\mp&\mathscr{E}^{1/3}e^{i2\pi/3}{\rm Ai}'\!\!\left(-\frac{E^{R\pm}}{\mathscr{E}^{2/3}}\,e^{i2\pi/3}\!\right)=0\\
\label{RobinEigenEquation1_D}
{\rm Ai}\!\left(-\frac{E^D}{\mathscr{E}^{2/3}}\,e^{i2\pi/3}\!\right)&=&0\\
\label{RobinEigenEquation1_N}
{\rm Ai}'\!\left(-\frac{E^N}{\mathscr{E}^{2/3}}\,e^{i2\pi/3}\!\right)&=&0.
\end{eqnarray}
\end{subequations}
From the last two equations it follows immediately that
\begin{eqnarray}
E_{res_n}^{\left\{_N^D\right\}}&=&-\left\{\begin{array}{c}a_n\\a_n'\end{array}\right\}\mathscr{E}^{2/3}e^{-i2\pi/3}\nonumber\\
\label{RobinEigenEnergiesDN1}
&=&\frac{1}{2}\left\{\begin{array}{c}a_n\\a_n'\end{array}\right\}\mathscr{E}^{2/3}\left(1+i3^{1/2}\right).
\end{eqnarray}
Eq.~\eqref{RobinEigenEnergiesDN1} means that the real parts of the resonance energies for the Dirichlet and Neumann BCs are always negative, while the corresponding half widths, as always in the method of zeroing the resolvent $\left(\hat{H}-E\right)^{-1}$, remain positive and both of them depend on the field as $\mathscr{E}^{2/3}$. For the Robin surface, the following asymptotes are taken from Eq.~\eqref{RobinEigenEquation1_R}:
\begin{subequations}\label{RobinEigenEnergiesAsymptot1}
\begin{eqnarray}\label{RobinEigenEnergiesAsymptot1SmallFields}
{E_{res}^{R\pm}}_n&=&\frac{1}{2}\,a_n\mathscr{E}^{2/3}\pm\mathscr{E}+i\frac{3^{1/2}}{2}\,a_n\mathscr{E}^{2/3},\quad\mathscr{E}\ll1\\
\label{RobinEigenEnergiesAsymptot1LargeFields}
{E_{res}^{R\pm}}_n&=&\frac{1}{2}\,a_n'\mathscr{E}^{2/3}\left(1\!\!+\!\!i3^{1/2}\right)\!\left(1\mp\!\frac{1}{a_n'^{\,\,2}}\frac{1}{\mathscr{E}^{1/3}}\right),\,\mathscr{E}\gg1.
\end{eqnarray}
\end{subequations}
A comparison of Eqs.~\eqref{RobinEigenEnergiesDN1}  and \eqref{RobinEigenEnergiesAsymptot1} manifests that at the weak fields the   Robin BC reduces basically to the Dirichlet one with a small admixture due to the interplay between the intensity $\mathscr{E}$ and non-zero extrapolation length $\Lambda$, while the high voltages essentially turn it into the Neumann surface where the higher order items, contrary to the previous limit, are the same for both the real and imaginary parts. Asymptotes from Eqs.~\eqref{RobinEigenEnergiesAsymptot1} present a general property of the interaction between the Robin BCs and the electric fields, which will be encountered below for the quasibound states as well as the true bound levels for the opposite direction of the field \cite{Olendski2}. It is also instructive to draw parallels with the similar states of the extremely localized potential whose energies at the weak fields are determined by Eq.~\eqref{DeltaComplexSolutionsTwoSets1_Set2}. It can be seen that in this limit they are almost identical to the difference, which is due to particle penetration to the left of the $\delta$ perturbation, being in the higher-order corrections. It is worth noting here that the Robin wall does not have the Gamow-Siegert resonances with the positive real parts of their energies since they are developed, as shown in the previous section, at $x<0$ where the motion for the present geometry is forbidden. In addition, the energy of the zero-field bound state at the small electric intensities is calculated as
\begin{equation}\label{RobinEigenEnergiesAsymptot2}
{E_{res}^{R-}}_0=-1-\frac{1}{2}\,\mathscr{E}-\frac{1}{8}\,\mathscr{E}^2-2i\exp\!\left(\!-\frac{4}{3}\frac{1}{\mathscr{E}}\right),\quad\mathscr{E}\ll1.
\end{equation}
The above formula shows that in this regime, due to the spatial asymmetry of the structure at $\mathscr{E}=0$, the real part of the energy is a linear function of the applied voltage while for the $\delta$-potential, which in the absence of the field is symmetric with respect to the transformation $x\rightarrow-x$, it depends quadratically on the electric intensity [see Eq.~\eqref{DeltaPotentialAsymptotics1}]. Note also a different pre-exponential factor in the expression for the half width in comparison to the attractive $\delta$-potential, Eq.~\eqref{DeltaHalfWidthAsymptotics1}:
\begin{equation}\label{RobinHalfWidthAsymptotics1}
{\Gamma_{res}^{R-}}_0=4\exp\!\left(\!-\frac{4}{3}\frac{1}{\mathscr{E}}\right),\quad\mathscr{E}\ll1.
\end{equation}
\begin{figure}
\centering
\includegraphics[width=\columnwidth]{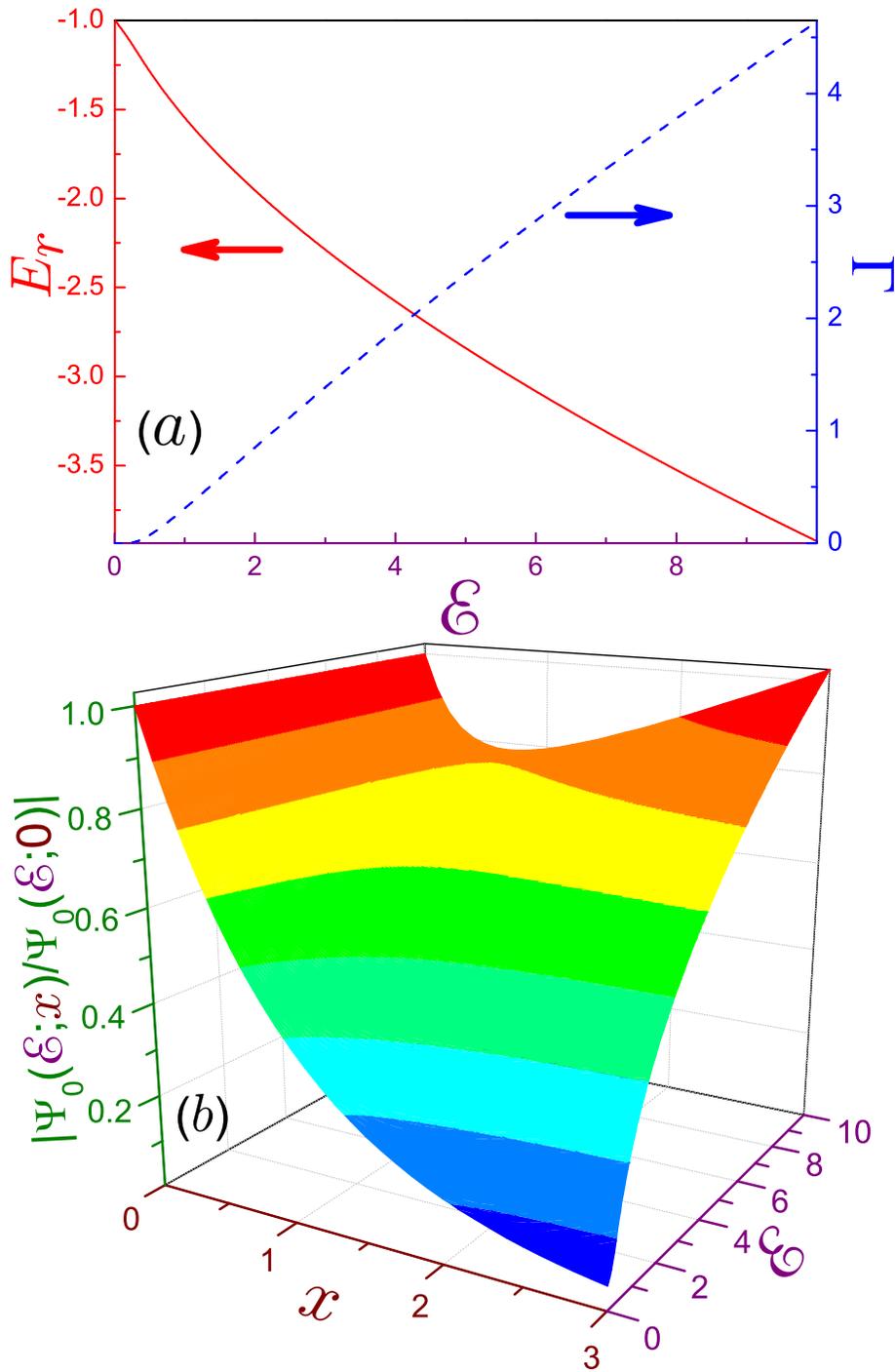}
\caption{\label{RobinComplexEnergyFunction0}
As Fig.~\ref{DeltaPotentialComplexEnergyFunction0} but for the Robin wall with negative extrapolation length.
}
\end{figure}

Fig.~\ref{RobinComplexEnergyFunction0} shows the evolution with the field of the resonance energy ${E_{res}^{R-}}_0$ and the associated waveform $\Psi_0(x)$. It can be seen that, contrary to the $\delta$-potential, the real part of the energy of the negative Robin wall in the whole range of the electric intensity is a decreasing function of $\mathscr{E}$: for low voltages, this decline is linear while for strong fields it asymptotically transforms to $a_1'\mathscr{E}^{2/3}/2$. The exponentially small increase of the half width at small fields is converted into the $3^{1/2}a_1'\mathscr{E}^{2/3}/2$ dependence at high $\mathscr{E}$. Of course, the 'exponential catastrophe' is an essential feature of the complex wave function of the Robin wall as well, as panel (b) of Fig.~\ref{RobinComplexEnergyFunction0} exemplifies. Its physical interpretation is the same as for the $\delta$-potential.

\begin{figure}
\centering
\includegraphics[width=\columnwidth]{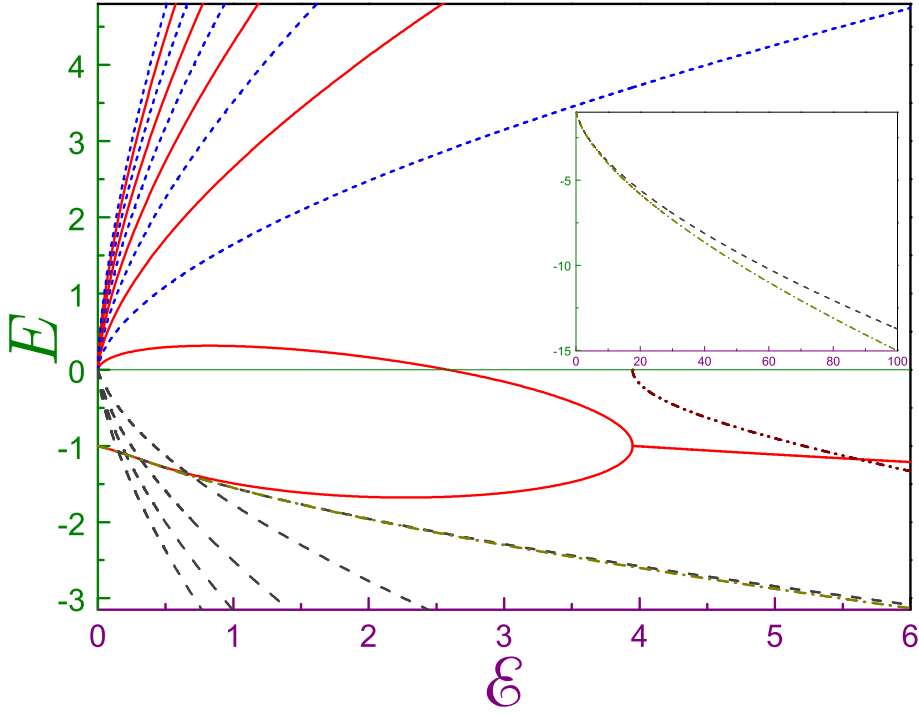}
\caption{\label{RobinEnergiesFig1}
Energy spectrum $E$ of the negative Robin wall as a function of the applied field $\mathscr{E}$ where the solid lines show REQB states that are solutions of Eq.~\eqref{RobinResonanceB1}, dotted curves depict energies satisfying Eq.~\eqref{RobinResonanceA1}, dashed lines are the Gamow-Siegert solutions from Eq.~\eqref{RobinEigenEquation1_R}, and the dash-dotted curve describes the evolution of the location of the maximum of the Wigner delay time $\tau_W^{R-}$, Eq.~\eqref{WignerRobin1_R}. The dash-dot-dotted line represents a negative imaginary component of the complex solution of Eq.~\eqref{RobinResonanceB1} at $\mathscr{E}\geq\mathscr{E}^{\times R-}$. The inset compares the complex-energy method and the location of the Wigner time maximum at the strong fields.
}
\end{figure}

To not deal with the divergences, the REQB states and delay time resonances have to be considered. Analysis of Eq.~\eqref{RobinScatteringMatrix1_R} reveals that for the Robin surface there exist two sets of quasibound levels that at low voltages are approximated by the Breit-Wigner formula. The first of these corresponds to the maximal distortion, $p^{R\pm}\left(\mathscr{E};E_n^{B\pm}\right)=1$, by the de Gennes interface of the wall-free function from Eq.~\eqref{ScatteringFunction0} when, as follows from Eq.~\eqref{RobinScatteringMatrix1_R}, the scattering matrix is a positive unity, $S^{R\pm}\left(\mathscr{E};E_n^{B\pm}\right)=1$, $n^\pm=\left\{\begin{array}{c}
1,2,\ldots\\
0,1,\ldots
\end{array}\right.$, and, accordingly, the total solution from Eq.~\eqref{ScatteringFunction1} degenerates to the Airy ${\rm Bi}$ function:
\begin{equation}\label{RobinFunctionB1}
\Psi_n^B(x)\sim {\rm Bi}\left(-\mathscr{E}^{1/3}x-\frac{E_n^B}{\mathscr{E}^{2/3}}\right).
\end{equation}
The corresponding energies $E_n^{B\pm}$ are real solutions of equation
\begin{equation}\label{RobinResonanceB1}
{\rm Bi}\!\left(-\frac{E}{\mathscr{E}^{2/3}}\right)\pm\mathscr{E}^{1/3}{\rm Bi}'\!\left(-\frac{E}{\mathscr{E}^{2/3}}\right)=0.
\end{equation}

The energies $E_n^{A\pm}$ of the second set of REQB states, which are found from equation
\begin{equation}\label{RobinResonanceA1}
{\rm Ai}\!\left(-\frac{E}{\mathscr{E}^{2/3}}\right)\pm\mathscr{E}^{1/3}{\rm Ai}'\!\left(-\frac{E}{\mathscr{E}^{2/3}}\right)=0,
\end{equation}
guarantee that the Robin wall does not disturb the free particle motion in the uniform electric field to its right: $p^{R\pm}\left(\mathscr{E};E_n^{A\pm}\right)=0$, $n=1,2,\ldots$. Note that Eq.~\eqref{RobinResonanceA1} describes the evolution of the true bound states for the opposite direction of the field \cite{Olendski2}. Corresponding Breit-Wigner half widths are calculated as:
\begin{subequations}\label{RobinHalfWidth1}
\begin{eqnarray}
\Gamma_n^{B\pm}(\mathscr{E})&=&\!-\!2\mathscr{E}\frac{{\rm Ai}\!\left(-E_n^B\!\left/\!\mathscr{E}^{2/3}\right.\!\!\right)\pm\mathscr{E}^{1/3}{\rm Ai}'\!\left(-E_n^B\!\left/\!\mathscr{E}^{2/3}\right.\!\!\right)}{\mathscr{E}^{1/3}{\rm Bi}'\!\left(-E_n^B\!\left/\!\mathscr{E}^{2/3}\right.\!\!\right)\mp E_n^B\,{\rm Bi}\!\left(-E_n^B\!\left/\!\mathscr{E}^{2/3}\right.\!\!\right)},\nonumber\\
\label{RobinHalfWidthB1}
&&n=\left\{\begin{array}{c}
1,2,\ldots\\
0,1,\ldots
\end{array}\right.\\
\Gamma_n^{A\pm}(\mathscr{E})&=&2\mathscr{E}\frac{{\rm Bi}\!\left(-E_n^A\!\left/\!\mathscr{E}^{2/3}\right.\!\!\right)\pm\mathscr{E}^{1/3}{\rm Bi}'\!\left(-E_n^A\!\left/\!\mathscr{E}^{2/3}\right.\!\!\right)}{\mathscr{E}^{1/3}{\rm Ai}'\!\left(-E_n^A\!\left/\!\mathscr{E}^{2/3}\right.\!\!\right)\mp E_n^A\,{\rm Ai}\!\left(-E_n^A\!\left/\!\mathscr{E}^{2/3}\right.\!\!\right)},\nonumber\\
\label{RobinHalfWidthA1}
&&n=1,2,\ldots.
\end{eqnarray}
\end{subequations}
In the limiting cases one has:

for the weak intensities, $\mathscr{E}\ll1$:
\begin{subequations}\label{AsymptotSmallFieldsScattering1}
\begin{align}\label{AsymptotSmallFieldsScatteringBE0}
E_0^{B-}&=-1-\frac{1}{2}\,\mathscr{E}-\frac{1}{8}\,\mathscr{E}^2
\intertext{(this equation under the transformation $\mathscr{E}\rightarrow-\mathscr{E}$ describes the lowest true bound state \cite{Olendski2})}
\label{AsymptotSmallFieldsScatteringBEn}
E_n^{B\pm}=&-b_n\mathscr{E}^{2/3}\pm\mathscr{E},\quad n=1,2,\ldots\\
\label{AsymptotSmallFieldsScatteringBG0}
\Gamma_0^{B-}&=4\exp\!\left(\!-\frac{4}{3}\frac{1}{\mathscr{E}}\right)\\
\label{AsymptotSmallFieldsScatteringBGn}
\Gamma_n^{B\pm}&=-2\frac{{\rm Ai}(b_n)}{{\rm Bi}'(b_n)}\mathscr{E}^{2/3}\\
\label{AsymptotSmallFieldsScatteringAEn}
E_n^{A\pm}&=-a_n\mathscr{E}^{2/3}\pm\mathscr{E}
\intertext{(comparing again with the reversed field configuration \cite{Olendski2}, it is important to stress that the only negative solution of Eq.~\eqref{RobinResonanceA1} is dropped for the present geometry, since it does not correspond to the free particle motion in the tilted potential)}
\label{AsymptotSmallFieldsScatteringAGn}
\Gamma_n^{A\pm}&=2\frac{{\rm Bi}(a_n)}{{\rm Ai}'(a_n)}\mathscr{E}^{2/3};
\end{align}
\end{subequations}

for the high voltages, $\mathscr{E}\gg1$:
\begin{subequations}\label{AsymptotLargeFieldsScattering1}
\begin{eqnarray}
\label{AsymptotLargeFieldsScatteringBEn}
E_n^{B\pm}&=&-b_n'\left(1\mp\frac{1}{b_n'^{\,\,2}\mathscr{E}^{1/3}}\right)\mathscr{E}^{2/3}\\
\label{AsymptotLargeFieldsScatteringBGn}
\Gamma_n^{B\pm}&=&-\frac{2}{b_n'}\frac{{\rm Ai}'(b_n')}{{\rm Bi}(b_n')}\left(1\pm\frac{1}{b_n'^{\,\,2}\mathscr{E}^{1/3}}\right)\mathscr{E}^{2/3}\\
\label{AsymptotLargeFieldsScatteringAEn}
E_n^{A\pm}&=&-a_n'\left(1\mp\frac{1}{a_n'^{\,\,2}\mathscr{E}^{1/3}}\right)\mathscr{E}^{2/3}\\
\label{AsymptotLargeFieldsScatteringAGn}
\Gamma_n^A&=&\frac{2}{a_n'}\frac{{\rm Bi}'(a_n')}{{\rm Ai}(a_n')}\left(1\pm\frac{1}{a_n'^{\,\,2}\mathscr{E}^{1/3}}\right)\mathscr{E}^{2/3}.
\end{eqnarray}
\end{subequations}
In the first limit, the outcomes of the two methods for the zero-field ground state are the same, as a comparison of Eqs.~\eqref{RobinEigenEnergiesAsymptot2} and \eqref{RobinHalfWidthAsymptotics1} with Eqs.~\eqref{AsymptotSmallFieldsScatteringBE0} and \eqref{AsymptotSmallFieldsScatteringBG0}, respectively, shows. The Gamow-Siegert states with the negative real components of their energies, which are described by Eqs.~\eqref{RobinEigenEnergiesAsymptot1}, do not have their quasibound counterparts, as was also the case for the $\delta$-potential. Next, we see once again that the asymptotes of the weak, Eqs.~\eqref{AsymptotSmallFieldsScatteringBEn} and \eqref{AsymptotSmallFieldsScatteringAEn} [strong, Eqs.~\eqref{AsymptotLargeFieldsScatteringBEn} and \eqref{AsymptotLargeFieldsScatteringAEn}], fields simplify the BC to the Dirichlet (Neumann) requirement with the small admixtures due to the tiny interaction between the electric intensity and the non-zero (finite) de Gennes distance. Moreover, all $A$ and non-zero $B$ states are characterized by negative half widths. This is seen from their expressions for large $n$:
\begin{subequations}\label{RobinAsymptotLargeN1}
\begin{eqnarray}\label{RobinAsymptotLargeN1_SmallField}
\Gamma_n^{\left\{_B^A\right\}}&=&-2\left[\frac{3\pi}{8}\left(4n\!-\!\left\{
\begin{array}{c}
1\\3
\end{array}
\right\}\right)\right]^{-1/3}\!\!\!\!\mathscr{E}^{2/3},\,\mathscr{E}\ll1,\,n\gg1\\
\label{RobinAsymptotLargeN1_LargeField}
\Gamma_n^{\left\{_B^A\right\}}&=&-2\left[\frac{3\pi}{8}\left(4n\!-\!\left\{
\begin{array}{c}
3\\1
\end{array}
\right\}\right)\right]^{-1/3}\!\!\!\!\mathscr{E}^{2/3},\,\mathscr{E}\gg1,\,n\gg1.
\end{eqnarray}
\end{subequations}
Negative half widths mean that for the Robin wall in the repulsive tilted potential, all field-induced REQB states are hole-like. It is easy to understand the absence of the electron-like excitations for the present geometry; namely, for the $\delta$-potential they are formed and modified in the area that lies on the opposite (when compared to the holes) side of the $x$ axis, but for the structure under consideration, the motion there is forbidden by the impenetrable Robin wall; accordingly, electron-like quasibound states have no room to be created and developed. This also means that with the hole being varied by the field, quasibound states have no partners with whom they can interact and therefore, in the framework of this model, they survive any electric intensity. The only exception is the lowest field-induced $B$ level, which interacts and ultimately collides with the state developed from the zero-field level. All these features are seen in Fig.~\ref{RobinEnergiesFig1}, which shows energies of the negative Robin wall as functions of the applied voltage. The zero-field level decreases its energy to the wide minimum of $E_{{B_0}_{min}}^{R-}=-1.678$ that is achieved at $\mathscr{E}=2.275$, after which it moves upwards towards the $B_1$ level, which has a broad maximum of ${E_{{B_1}_{max}}^{R-}}=0.314$ at $\mathscr{E}=0.825$ and crosses zero at $\mathscr{E}_{E=0}^{R-}=\mathscr{E}_f$; this is easily derived from Eq.~\eqref{RobinResonanceB1}. At the merger, and in addition to this equation, the derivative with respect to the energy of the left-hand side should change to zero, which means that the energy at the amalgamation is $E^{R_-^\times}=-1$ while the corresponding voltage $\mathscr{E}^{R_-^\times}=3.94827\ldots$ is found numerically from the following equation
\begin{equation}\label{RobinMergerField1}
{\rm Bi}\!\left(\mathscr{E}^{-2/3}\right)-\mathscr{E}^{1/3}{\rm Bi}'\!\left(\mathscr{E}^{-2/3}\right)=0.
\end{equation}
Thus, even though the lowest field-induced quasibound state of the de Gennes wall turns to zero at the same electric intensity as its $\delta$-well counterpart, the coalescence of the levels takes place at higher voltages, which can be interpreted as the stronger binding of the particles by the negative Robin surface. Close to amalgamation, the energies of the merging levels are:
\begin{equation}\label{AsymptotRobinCriticalField1}
E_{\left\{_{B_0}^{B_1}\right\}}=-1\pm\left[\frac{2}{3}\left(\mathscr{E}^{R_-^\times}-\mathscr{E}\right)\right]^{1/2},\quad\mathscr{E}\rightarrow\mathscr{E}^{R_-^\times}.
\end{equation}
Beyond the coalescence, the system again exhibits two complex conjugate solutions whose evolution with the voltage is also shown in Fig.~\ref{RobinEnergiesFig1}. Their existence is another implication of the electron-hole coupling and interaction at the strong fields. The energies of all other higher-lying quasibound states monotonically increase in the whole range of the voltage change from the Dirichlet-like dependence at the small $\mathscr{E}$, Eqs.~\eqref{AsymptotSmallFieldsScatteringBEn} and \eqref{AsymptotSmallFieldsScatteringAEn}, to the almost Neumann BC, Eqs.~\eqref{AsymptotLargeFieldsScatteringBEn} and \eqref{AsymptotLargeFieldsScatteringAEn}, at large electric intensities. As mentioned above, these positive energy levels, contrary to the $\delta$-well, do not have their Gamow counterparts. Several negative real parts of the field-induced complex energies are also plotted in the figure. Qualitatively, their behavior is similar to the $\delta$-geometry. In particular, all of them cross the level evolved from the zero-field bound state whose energy is a monotonically decreasing function of the field; namely, at low voltages it is a negative linear function of the electric intensity according to Eq.~\eqref{RobinEigenEnergiesAsymptot2} while at strong $\mathscr{E}$ it obeys the dependence $a_1'\mathscr{E}^{2/3}/2$ from Eq.~\eqref{RobinEigenEnergiesAsymptot1LargeFields} with $n=1$.

\begin{figure}
\centering
\includegraphics[width=\columnwidth]{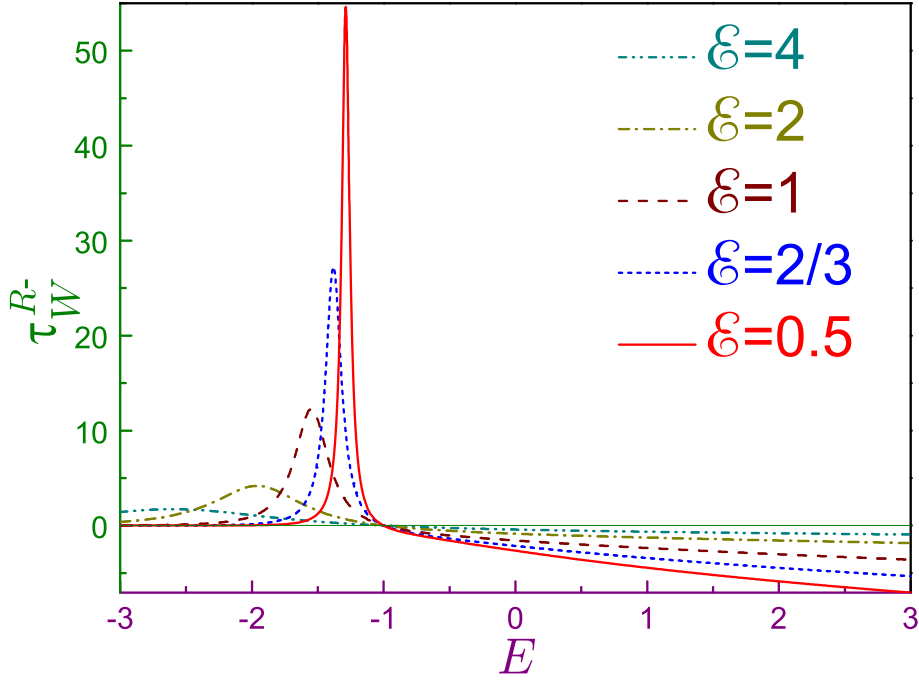}
\caption{\label{RobinWignerTime}
Wigner delay time $\tau_W^{R-}$ of the negative Robin wall as a function of the energy $E$ for several electric fields where solid line is for $\mathscr{E}=0.5$, dotted line is for $\mathscr{E}=2/3$, dashed curve -- for $\mathscr{E}=1$, dash-dotted line -- for $\mathscr{E}=2$, and dash-dot-dotted one -- for $\mathscr{E}=4$. All curves intersect at $(-1,0)$, as follows from Eq.~\eqref{WignerRobin1_R}.
}
\end{figure}

For the delay time, explicit evaluation yields:
\begin{subequations}\label{WignerRobin1}
\begin{eqnarray}
&&\tau_W^{R\pm}(\mathscr{E};E)=-\frac{2}{\pi\mathscr{E}^{2/3}}\times\nonumber\\
\label{WignerRobin1_R}
&&\frac{E+1}{{\rm Ai}_0^2\!+\!\!{\rm Bi}_0^2\pm2\mathscr{E}^{1/3}\!\left({\rm Ai}_0{\rm Ai}_0'\!+\!\!{\rm Bi}_0{\rm Bi}_0'\right)\!+\!\!\mathscr{E}^{2/3}\!\left({{\rm Ai}_0'}^2\!\!+{{\rm Bi}_0'}^2\!\right)}\\
\label{WignerRobin1_D}
&&\tau_W^D(\mathscr{E};E)=-\frac{2}{\pi\mathscr{E}^{2/3}}\frac{1}{{\rm Ai}_0^2+{\rm Bi}_0^2}\\
\label{WignerRobin1_N}
&&\tau_W^N(\mathscr{E};E)=-\frac{2}{\pi\mathscr{E}^{4/3}}\frac{E}{{{\rm Ai}_0'}^2+{{\rm Bi}_0'}^2},
\end{eqnarray}
\end{subequations}
where the same convention regarding the subscript '$0$' in Eq.~\eqref{DeltaTimeDelay1} is used. It can be seen that the Dirichlet delay time is always negative, which means that the corresponding wall cannot capture the particle for a prolonged period but rather is in a hurry at any $E$ to reflect it back, whereas its Neumann counterpart bears the opposite sign to that of the energy. The delay time decays exponentially for large negative energies
\begin{subequations}\label{WignerRobinAsymptot1}
\begin{align}
\tau_W^{R\pm}(\mathscr{E};E)&=\tau_W^N(\mathscr{E};E)=-\tau_W^D(\mathscr{E};E)\nonumber\\
\label{WignerRobinAsymptot1_Negative}
&=2\frac{|E|^{1/2}}{\mathscr{E}}\exp\!\left(-\frac{4}{3}\frac{|E|^{3/2}}{\mathscr{E}}\right),\quad E\ll-1,\\
\intertext{while in the opposite case it is transformed into the BC-independent smoothly varying negative quantity:}
\tau_W^{R\pm}(\mathscr{E};E)&=\tau_W^N(\mathscr{E};E)=\tau_W^D(\mathscr{E};E)\nonumber\\
\label{WignerRobinAsymptot1_Positive}
&=-2\frac{E^{1/2}}{\mathscr{E}},\quad E\gg1.
\end{align}
\end{subequations}

Observe that the leading term of the first limit coincides with the one from the corresponding expression for the $\delta$-potential, Eq.~\eqref{DeltaTimeDelayAsymptotics1_NegativeEnergy}. However, for large positive energies the dependencies of the two geometries are very different, since the Robin wall forbids motion in the area $x<0$ where the $\delta$-well resonances described by Eqs.~\eqref{AsymptotDeltaStrongField1}, \eqref{DeltaTimeDelayAsymptotics3} and \eqref{DeltaResonanceHalfWidth1} are formed. Note that the absolute value of the right-hand side of Eq.~\eqref{WignerRobinAsymptot1_Positive} is just the time needed for the classical particle to travel from the Robin surface to the wall-free quasi-classical turning point $x_{qc}=-E/\mathscr{E}$ and return at $x=0$.

Fig.~\ref{RobinWignerTime} depicts the Wigner delay time as a function of energy for several voltages. Field-dependent asymptotes from Eqs.~\eqref{WignerRobinAsymptot1} are clearly seen. Another feature that follows from Eq.~\eqref{WignerRobin1_R} and is shown in the plot is the fact that $\tau_W^{R-}$ at arbitrary electric intensities vanishes if the energy is a negative unity, $E=-1$. As the figure depicts, the asymmetric structure of the attractive Robin surface is characterized by one resonance only, which stems from the zero-field bound state: at very small $\mathscr{E}$ it is located at the energies from Eq.~\eqref{AsymptotSmallFieldsScatteringBE0} with the narrow FWHM coinciding with the half width from Eq.~\eqref{AsymptotSmallFieldsScatteringBG0} while its peak value in this regime is
\begin{equation}\label{WignerLimitMax1}
{\tau_W^{R-}}_{\!\!max}(\mathscr{E})=\exp\left(\frac{4}{3}\frac{1}{\mathscr{E}}\right),\quad\mathscr{E}\rightarrow0.
\end{equation}
The growing field decreases the maximum and widens the FWHM of the resonance. A change in its location with the applied voltage is shown by the dash-dotted line in Fig.~\ref{RobinEnergiesFig1}. At small electric intensities, $\mathscr{E}\ll1$, the three energies describing the transformation of the  zero-field bound state by the three different methods are practically equal to each other. At $\mathscr{E}\gtrsim0.7$ the quasibound state starts to deviate upwards from the other two energies, which, contrary to the symmetric $\delta$-structure, monotonically decrease across the whole range of the increasing voltage. Another difference between the two geometries is the fact that for the Robin interface the Gamow energy lies above its Wigner counterpart.

\begin{figure}
\centering
\includegraphics[width=\columnwidth]{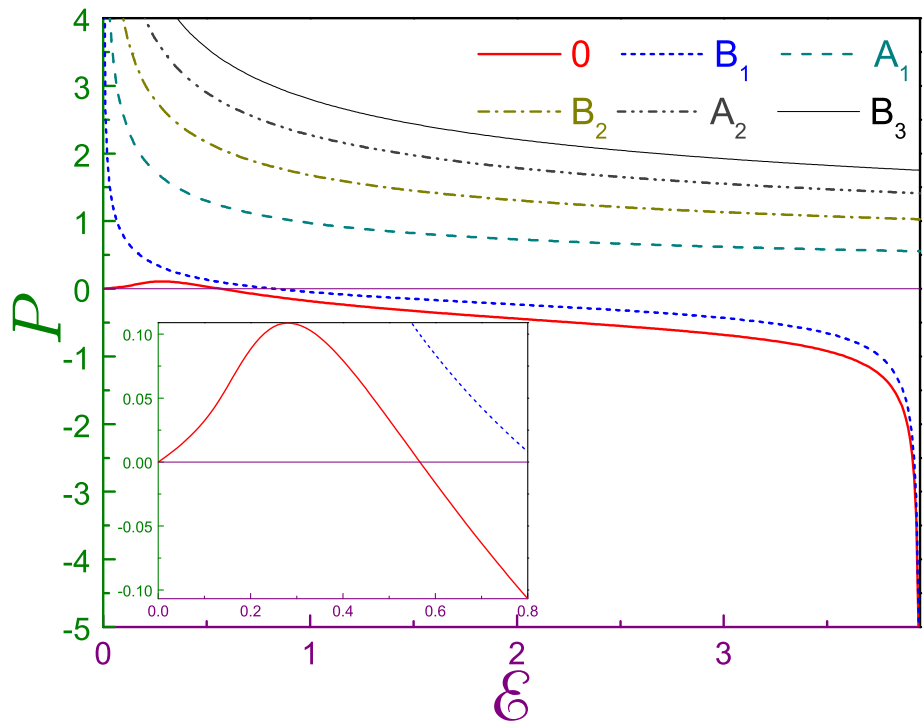}
\caption{\label{RobinPolarizationFig1}
Dipole moments $P^{R-}$ of the quasibound states for the negative Robin wall as a function of the electric field $\mathscr{E}$. 
The same line convention as in Fig.~\ref{DeltaPolarizationFig1} is used. Inset shows the enlarged view at low voltages.}
\end{figure}

As the considered structure is asymmetric, the zero-field mean coordinate $x$ needs to be evaluated to correctly calculate the polarization. For the ground state, an elementary computation where the integrand contains the square of the function $\Psi$ from Eq.~\eqref{FunctionRobinZeroFields1} produces $\langle x\rangle_{\mathscr{E}=0}=1/2$. Then, at low voltages, this term is exactly compensated by the linear contribution to the energy, as an application of the Hellmann-Feynman theorem, Eq.~\eqref{Polarization3}, to Eq.~\eqref{AsymptotSmallFieldsScatteringBE0} shows:
\begin{equation}\label{RobinPolarizationSmallFields1}
P_0^{R-}(\mathscr{E})=\frac{1}{4}\mathscr{E},\quad\mathscr{E}\ll1.
\end{equation}
Note the different slope of this polarization compared to the $\delta$-well mean coordinate, Eq.~\eqref{DeltaPolarization3}. Eq.~\eqref{RobinPolarizationSmallFields1} remains the same for the opposite direction of the field \cite{Olendski2}.  In turn, for the excited REQB states, a calculation of the zero-field $\langle x\rangle$ has to be performed at $E=0$ since the energies of all these levels approach zero in the limit of the vanishing electric intensities; see Eqs.~\eqref{AsymptotSmallFieldsScatteringBEn} and \eqref{AsymptotSmallFieldsScatteringAEn} and Fig.~\ref{RobinEnergiesFig1}. Then, the solution of the corresponding Schr\"{o}dinger equation, Eq.~\eqref{Schrodinger3}, with $\mathscr{E}=E=0$ that satisfies the BC from Eq.~\eqref{BCRobin2}, reads:
\begin{equation}\label{FakeSolution1}
\Psi_{n\geq1}^{R\pm}(x)\sim x\pm1\quad {\rm at}\quad\mathscr{E}=E=0.
\end{equation}
This waveform has to be discarded since it diverges at infinity. The only remaining trivial solution $\Psi=0$ means that $\langle x\rangle_{\mathscr{E}=0}=0$ for $n\geq1$. Accordingly, the expression for the dipole moment reads for, e.g., the $B$ levels:
\begin{equation}\label{RobinPolarization1}
P_{B_n}^{R-}=\frac{1}{3}\frac{2\frac{E^2}{\mathscr{E}}{\rm Bi}\!\left(-\frac{E}{\mathscr{E}^{2/3}}\right)\!+\!\mathscr{E}^{1/3}\left(2\frac{E}{\mathscr{E}}-\!\!1\right){\rm Bi}'\!\left(-\frac{E}{\mathscr{E}^{2/3}}\right)}{E\,{\rm Bi}\!\left(-\frac{E}{\mathscr{E}^{2/3}}\right)+\mathscr{E}^{1/3}{\rm Bi}'\!\left(-\frac{E}{\mathscr{E}^{2/3}}\right)},\,n\geq1.
\end{equation}
The limiting cases, in addition to Eq.~\eqref{RobinPolarizationSmallFields1}, are:
\begin{subequations}\label{RobinPolarizationAsymptote1}
\begin{eqnarray}\label{RobinPolarizationAsymptote1_ABsmall1}
P_{\left\{_{B_n}^{A_n}\right\}}^{R-}&=&-\frac{2}{3}\left\{\!\!\begin{array}{c}
a_n\\b_n
\end{array}\!\!\right\}\frac{1}{\mathscr{E}^{1/3}},\quad n\geq1,\,\mathscr{E}\ll1\\
\label{RobinPolarizationAsymptote1_ABlarge1}
P_{\left\{_{B_n}^{A_n}\right\}}^{R-}&=&-\frac{2}{3}\left\{\!\!\begin{array}{c}
a_n'\\b_n'
\end{array}\!\!\right\}\frac{1}{\mathscr{E}^{1/3}},\quad\left\{\!\begin{array}{c}
n\geq1\\n\geq2
\end{array}\!\right\},\,\mathscr{E}\gg1\\
\label{RobinPolarizationAsymptote1_Bcoalesc1}
P_{B_n}^{R-}&=&\!\!-\!\!\left[6\left(\mathscr{E}^{R_-^\times}\!\!-\!\!\mathscr{E}\right)\right]^{-1/2}\!\!\!-\frac{1}{2}\delta_{n0},\,n=0,1,\,\mathscr{E}\!\!\rightarrow\!\!\mathscr{E}^{R_-^\times},
\end{eqnarray}
\end{subequations}
where $\delta_{nm}$ is the Kronnecker $\delta$.

\begin{figure}
\centering
\includegraphics[width=0.7\columnwidth]{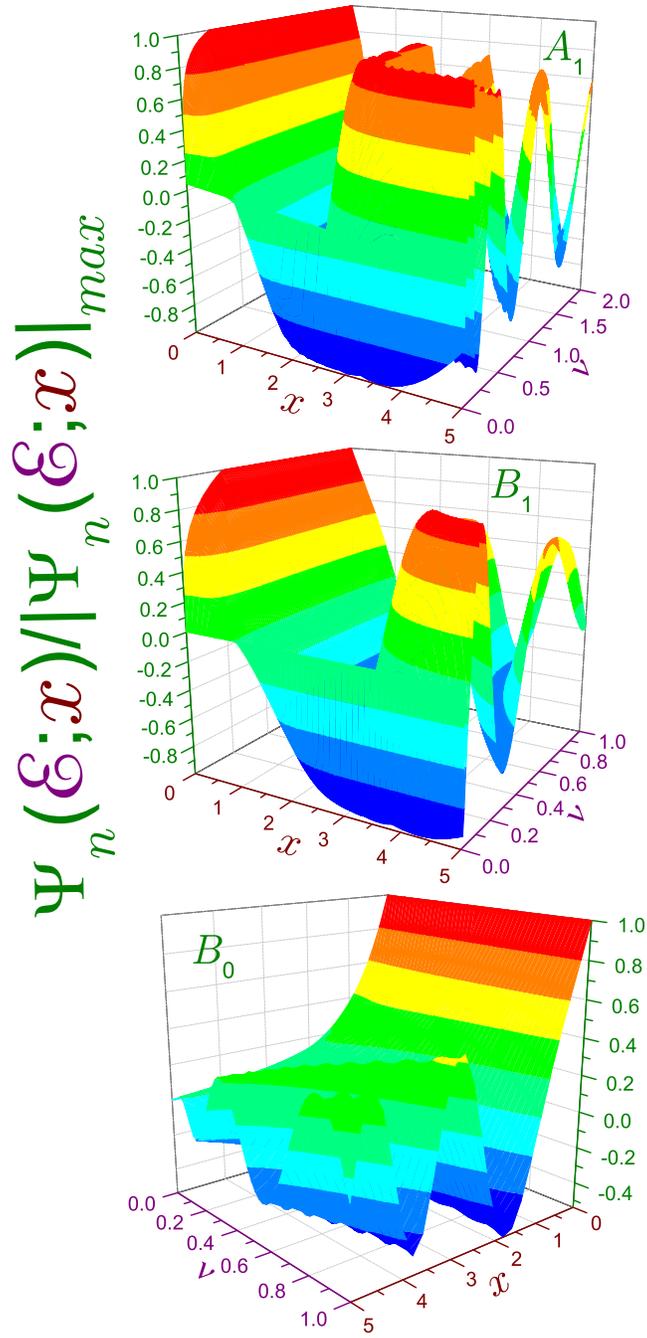}
\caption{\label{RobinFunctionsFig1}
Wave functions $\Psi(\mathscr{E};x)$ of the first three lowest states of the attractive Robin wall, normalized to the maximum of their absolute values in terms of the coordinate $x$ and normalized electric field $\nu\equiv\mathscr{E}/\mathscr{E}^{R_-^\times}$. Note the different ranges of the $\nu$ axis of the upper panel compared to those for $B_0$ and $B_1$. The lower limit of the vertical axis for the $B_0$ state also differs from its counterparts for $B_1$ and $A_1$ levels.}
\end{figure}

Fig.~\ref{RobinPolarizationFig1} shows several polarizations for the attractive surface in the electric field. The qualitative behavior of the dipole moments of the two lowest states is similar to their counterparts considered in Sec.~\ref{Sec_Delta}, with the quantitative differences, one of which was discussed in the previous paragraph, being due to the inability of the particle to penetrate into the area to the left of the wall; namely, the initial linear growth of the ground-state polarization with the field is followed by its maximum of ${P_0^{R-}}_{\!\!\!\!\!max}=0.1087$, which is achieved at $\mathscr{E}=0.281$. As a result of a subsequent decrease, the dipole moment returns to zero at $\mathscr{E}_{P_0=0}^{R-}=0.567$ compared to $\mathscr{E}_{P_0=0}^{\delta-}=0.739$ for the $\delta$-well. The turnaround behavior of the ground state polarization is caused by its interaction with the neighboring hole-like level, with its dipole moment decreasing from the infinitely large values at intensities $\mathscr{E}$ close to zero, which correspond to its location far to the right, to the large negative magnitudes when it approaches the coalescence field $\mathscr{E}^{R_-^\times}$. At the amalgamation, the two waveforms coincide. Their evolution with the field is shown in the two lowest panels of Fig.~\ref{RobinFunctionsFig1}. Higher-lying hole-like states, contrary to the $B_1$ level, do not have counterparts with whom they collide at increasing voltage; accordingly, their polarizations monotonically decrease to zero as the electric intensity grows and this corresponds to the squeezing of the wave functions closer to the wall. This shift to the left is depicted in the upper plot of  Fig.~\ref{RobinFunctionsFig1}.

\section{Concluding Remarks}\label{Conclusions}
To correctly describe a physical phenomenon, scientists need to use a proper mathematical model. Here, to overcome a definite discord in the existing literature, an attempt was made to identify and analyze three different processes taking place in 1D quantum nanostructures in the uniform electric field. Mathematically, they are described by the distinct properties of the corresponding scattering matrix; namely, its poles, real values, and zeros of the second derivative of its phase. Using the examples of the 1D $\delta$-potential and Robin wall, it was demonstrated that the zero-resolvent method produces mathematically correct solutions in the form of the {\em complex} eigen energies $E$ and eigen functions $\Psi(x)$ that, at $\mathscr{E}\neq0$, show a divergence at large distances [see panels (b) of Figs.~\ref{DeltaPotentialComplexEnergyFunction0} and \ref{RobinComplexEnergyFunction0}]. This is referred to as the 'exponential catastrophe' \cite{Bohm1}, which can, however, be cured by the time-dependent interpretation. The non-zero current density for this model denotes a leakage of the electron away from the QW. For the real energies, the total net current is zero, and conditions of the formation of the quasibound states are formulated as a requirement of the largest distortion by the potential of the free particle motion in the electric field. Analysis of this model reveals that the weak field splits the positive-energy continuum into electron- and hole-like quasibound levels that, with increasing voltage, strongly interact between themselves and ultimately collide with each other at the breakdown voltage. At even stronger electric intensities, the corresponding equation possesses two complex-conjugate solutions that may correspond to the formation of the composite exciton-like structure, where the motion of one part is correlated with a second constituent. Amalgamation of the levels is accompanied by the divergence of the associated dipole moments. The total number of each kind of levels is specified by the zero-field geometry; for example, the $\delta$-potential, which is symmetric with respect to the inversion $x\rightarrow-x$, has a countably infinite number of both electron and hole excitations while the asymmetric Robin wall, which forbids motion to its left, is characterized by the only electron state developed from the zero-field bound level, in addition to the infinite set of hole orbitals.

Another set of resonances is associated with the maxima of the Wigner delay time $\tau_W$, which is a derivative of the phase $\varphi_S$ of the scattering matrix with respect to energy. The number of these extrema is again determined by the zero-field symmetry of the structure and is infinite for the $\delta$-potential and just one for the asymmetric attractive Robin wall. It was shown that at low voltages, the results of the calculation of the resonances and quasibound states developed from those of the field-free geometry are identical, but they diverge from each other for stronger electric intensities. Other similarities and differences between the three models were also discussed.

Having theoretically seen the peculiarities of the electric field effect, it is natural to wonder whether they could be experimentally verified. First, let us consider the atomic system; namely, the hydrogen ion ${\rm H}^-$ near the-one electron threshold \cite{Stewart1}. Of course, modeling a 3D structure with a 1D attractive potential is a rather crude approximation \cite{Emmanouilidou2}, but we are only interested in estimating the orders of magnitude. The reported binding energy $E=-0.7542$ eV \cite{Stewart1} corresponds to the very deep $\delta$ well with length $\Lambda=-2.2\times10^{-15}$ m [see Eq.~\eqref{EnergyDeltaZeroFields1}]. Accordingly, converting the dimensionless value of the fundamental dissociation field from Eq.~\eqref{FundamentalField1} into regular, normalized units, static electric intensities $\sim10^{16}$ V/m ($\sim10^5$ a.u.) that far exceed the experimentally available voltages of $\sim10^8$ V/m ($\sim10^{-3}$ a.u.) are gained \cite{Stewart1}. Thus, the behavior near the breakdown point $\mathscr{E}_f$ is hardly verifiable for atomic systems. The situation, however, changes dramatically for man-made semiconductor structures. Impressive advances in nanotechnology over the last two decades have allowed the development of low-dimensional artificial patterns of almost any desirable shape. From this point of view, it should be noted that early conjectures \cite{Pazma1,Fulop1} regarding the possibility of mimicking the Robin wall with $\Lambda<0$ by the limit of finite regularized potentials were corroborated by consideration of the following term $V(x)$ in Eq.~\eqref{Hamiltonian1} at $\mathscr{E}=0$:
\begin{equation}\label{Potential2}
V(x)=\left\{\begin{array}{cc}
\infty,&x<0\\
-V_0,&0<x<d\\
0,&d<x<\infty
\end{array}\right.
\end{equation}
with positive $V_0$ and $d$ \cite{Olendski6}. It was shown that at
\begin{equation}\label{Potential4}
\frac{1}{8}\frac{\pi^2\hbar^2}{md^2}<V_0<\frac{9}{8}\frac{\pi^2\hbar^2}{md^2}
\end{equation}
only one negative-energy state does exist (which is a necessary characteristic of the negative de Gennes surface) with its depth being controlled by the interrelation between $V_0$ and $d$. Accordingly, by applying to it appropriately directed voltage, it is possible to reach the dissociation field at any desired intensity.

Idealized models of the $\delta$-potential and Robin wall allow us to gain quite simple analytical results that facilitate their physical interpretation. A more realistic approach should take into account the finite width $d$ and depth $V_0$ of the structures resulting, in particular, in several zero-field bound states, a number of which depends on the ratio $V_0d^2$. Contradictory \cite{Austin4} complex-Airy-function \cite{Ahn1,Emmanouilidou1} and real energy \cite{Austin2} calculations have been carried out for the symmetric finite QW. Another possibility to get at least two bound states for the voltage-free configuration is to use more than one $\delta$-potential. Analysis of the location of the poles in the $E$ plane revealed the dependencies of the {\em complex} energy resonances on the field and the separation between the two extremely localized wells \cite{Korsch1,Alvarez2}. For the same system, even a superficial analysis based on the {\em real} Airy functions revealed a very complicated structure for the positive energy part of the spectrum \cite{Glasser1,Carpena1}. Relevant to our research, we mention here in passing that at $\mathscr{E}=0$, the Robin BCs are equivalent to a pair of 1D Dirac $\delta(x)$-$\delta'(x)$ interactions for a couple of critical values of the $\delta'$ coupling \cite{Kurasov1,Gadella1}. In light of the results presented above, it does make sense to readdress the problem of the two extremely localized QWs or barriers in the electric field in the same way as was carried out above for the single $\delta$-potential.

\section{Acknowledgement}\label{sec_6}
Constructive comments of anonymous referees are gratefully acknowledged.

\end{document}